\documentclass[preprint,12pt]{aastex}

\shorttitle{The Overdensity in Virgo and Sagittarius Debris}
\shortauthors{Newberg, Yanny et al.}

\begin{document}

\title{The Overdensity in Virgo, Sagittarius Debris, and the Asymmetric Spheroid}

\author{
Heidi Jo Newberg\altaffilmark{\ref{RPI}},
Brian Yanny\altaffilmark{\ref{FNAL}},
Nate Cole\altaffilmark{\ref{RPI}},
Timothy C. Beers\altaffilmark{\ref{msu}},
Paola Re Fiorentin\altaffilmark{\ref{MPH}},
Donald P. Schneider\altaffilmark{\ref{PSU}},
Ron Wilhelm\altaffilmark{\ref{TTech}}
}

\altaffiltext{1}{Dept. of Physics, Applied Physics and Astronomy, Rensselaer
Polytechnic Institute Troy, NY 12180; heidi@rpi.edu\label{RPI}}

\altaffiltext{2}{Fermi National Accelerator Laboratory, P.O. Box 500, Batavia,
IL 60510; yanny@fnal.gov\label{FNAL}}

\altaffiltext{3}{Department of Physics and Astronomy, Center for the Study of Cosmic Evolution, and Joint Institute for Nuclear Astrophysics, Michigan State University, East Lansing, MI 48824\label{msu}}

\altaffiltext{4}{Max-Planck-Institute f\"ur Astronomy, K\"onigstuhl 17, D-69117 Heidelberg, Germany \label{MPH}}

\altaffiltext{5}{Department of Astronomy and Astrophysics, The Pennsylvania State University, University Park, PA 16802\label{PSU}}

\altaffiltext{6}{Department of Physics, Texas Tech University, Lubbock, TX \label{TTech}}

\begin{abstract}

We investigate the relationship between several previously identified Galactic halo stellar 
structures in the direction of Virgo using imaging and spectroscopic observations 
of F turnoff stars and blue horizontal branch stars
from the Sloan Digital Sky Survey (SDSS) and the Sloan Extension for Galactic 
Understanding and Exploration (SEGUE).  
We show that the Sagittarius dwarf leading tidal tail 
does not pass through the solar neighborhood; it misses the Sun by more than
15 kpc, passing through the Galactic plane outside the Solar Circle. It also
is not spatially coincident with the large stellar overdensity S297+63-20.5 in
the Virgo constellation.  S297+63-20.5 has a distinct turnoff 
color and kinematics.  Faint ($g_0\sim 20.3$) turnoff stars 
in S297+63-20.5 have line-of-sight, Galactic standard of rest 
velocities $V_{gsr}=130\pm 10 \>\rm km\>s^{-1}$, opposite in sign 
to infalling Sgr tail stars.  The path of the Sgr leading tidal tail
is also inconsistent with the positions of some of the nearer stars with
which it has been associated, and whose velocities have favored
models with prolate Milky Way potentials.  We additionally show that the 
number densities of brighter ($g_0 \sim 19.8$) F turnoff stars are 
not symmetric about the Galactic center, and that 
this discrepancy is not primarily due to the S297+63-20.5 moving group.  
Either the spheroid is asymmetric about the Galactic center,
or there are additional substructures that 
conspire to be on the same side of the Galaxy as S297+63-20.5.
The S297+63-20.5 overdensity in Virgo is likely
associated with two other previously identified Virgo substructures: the
Virgo Stellar Stream (VSS) and the Virgo Overdensity (VOD).  However, 
the velocity difference between the VSS and S297+63-20.5 and the
difference in distance estimates between the VOD and S297+63-20.5 must be reconciled.
\end{abstract}


\section{Introduction\label{intro}}

A number of studies have noted that there are more spheroid stars in the Virgo constellation
than were expected from smooth spatial models of the spheroid star density.  The first 
indication of spheroid substructure in this region of the sky came
from the \citet{vetal01} discovery of five excess RR Lyrae stars from Quasar Equatorial
Survey Team (QUEST) survey data.  Although
a small number, it was a significant overdensity in a small area.  The QUEST collaboration later
identified an excess of 23 RR Lyrae variables with this clump.  These RR Lyrae
stars have apparent magnitudes $16.5<V_0<17.5$, and are within the Galactic coordinates
$279^\circ<l<317^\circ$, $60^\circ<b<63^\circ$.
The collaboration later dubbed this feature the ``$12^{\rm h}.4$ Clump" \citep{zvgw04}.  
Subsequent spectra of the
RR Lyraes and BHB stars in this clump \citep{detal06} revealed a moving group at (RA, DEC) 
$=(186^\circ,-1^\circ)$ with a line-of-sight,
Galactic standard of rest velocity of $V_{gsr}=99.8\rm \>km\>s^{-1}$ and a distance of 19
kpc from the Sun.  They suggest that this moving group
be called the Virgo Stellar Stream (VSS).

The existence of an overdensity of stars in the stellar spheroid near 
$(l,b)=(297^\circ,63^\circ)$ was 
independently confirmed by \citet{nyetal02} using measurements of blue turnoff stars from the Sloan 
Digital Sky Survey (SDSS; York et al. 2000), and identified as S297+63-20.5.\footnote{In \citet{nyetal02} this structure
was incorrectly labeled 297+63-20.  The ``S" signifies stream (or substructure), followed by the Galactic longitude, then
Galactic latitude, and the apparent magnitude of the reddening corrected $g_0$ turnoff.  
In the present paper we correct the name to be S297+63-20.5; both names will refer to 
the same overdensity of stars.}
  The estimated distance of this structure
from the Sun, calculated from $g_0 = 20.5$ turnoff stars, is $r_{\rm Sun} = 18$ kpc, but the 
color-magnitude diagram of this region does not have a tight main 
sequence, indicating that the structure is likely to be dispersed in distance.  
This overdensity appears to match the position of the ``$12^{\rm h}.4$ Clump", but is somewhat
offset from the VSS, which is at $(l,b)=(288^\circ,62^\circ)$.  

Because the mapping between the physical structure on the sky and
its nomenclature in the literature for the apparent excess in the number of
spheroid stars in the Virgo constellation is unclear,
we will refrain for now from identifying the S297+63-20.5 stars with the VSS, 
though they are at about the same distance and position in the sky.
We will return to the question of which named spheroid structures are related to each other
in the discussion section of this paper.

\citet{mswo03} also identify a structure in Virgo using the stellar catalogs from the 2MASS 
survey, and connect it to ``M giants tens of kiloparsecs above the Galactic plane that are 
in the heart of the descending, foreshortened northern loop," which is their 
terminology for the leading tidal tail of the Sgr dwarf galaxy.  Any overdensity in this 
part of the sky is close to the orbital plane of the Sgr dwarf galaxy, and might be related to 
the Sgr dwarf tidal debris.  Later papers, for example \citet{ljm05}, show both the 
leading and trailing tidal tails of the Sgr dwarf looping around through a similar 
position on the sky, near the identified overdensities in Virgo, before passing through
the Galactic plane very close to the Sun.  Since the Sagittarius tidal 
tails are quite prominent in the spheroid, and the observed overdensity or 
overdensities in the Virgo region are near the orbital plane of the Sagittarius dwarf, a 
connection between these structures is worthy of exploration.

\citet{jetal06} identify a large overdensity (the Virgo Overdensity, VOD) in the Virgo 
constellation, which may or may not be related to the previously named substructures.  Using 
photometric parallaxes for a large number of SDSS stars, they show that this large overdensity
is $\approx 5-15$ kpc above the plane and covers $1000$ sq. deg. of sky.  They detect no 
downturn in the star counts toward lower Galactic latitudes, indicating that the structure 
could continue further into the Galactic plane than the SDSS observations probe.  At that time
the SDSS had not yet probed Galactic latitudes lower than about $60^\circ$ near $l=300^\circ$.  
\citet{jetal06} interpret the overdensity as a significant spheroid substructure that might be an
invading dwarf galaxy.  The estimated distance to the VOD is smaller than the distance to the
VSS or S297+63-20.5.  It is unclear whether the nearer distance is real, or if distance
uncertainties are large enough that the VOD, VSS, and S297+63-20.5 could describe the same
physical structure.  In \S 2 and \S 3 of this paper we will discuss S297+63-20.5 only; the relationship
of this overdensity to the VOD and VSS will be deferred to the discussion.

More recently, \citet{mpjai06} suggest that the invading dwarf galaxy might be the Sgr 
dwarf galaxy leading tidal tail, and show that the \citet{ljm05} model passes through the observed 
location of the VOD, and that the stellar density of the VOD is similar to the model
predictions.  They suggest that if the Sgr dwarf tidal stream 
passes through the solar neighborhood, it will be detected as an excess of stars with 
negative radial velocities in the northern Galactic hemisphere.  This paper treats the 
VOD as a separate entity from the VSS, but located in a similar position in space.  The 
VOD is described by these authors as likely to be associated with the Sgr leading arm, and 
therefore infalling, and the VSS is a group of stars nearly coincident in sky position, but 
with a different vertical and kinematic structure.

Because the Sgr dwarf galaxy tidal tails can help constrain the shape of gravitational 
potential of the Milky Way and its dark matter halo, they are the topic of 
considerable study, independent from their possible connection with the observed 
overdensity in Virgo.  One can in principle, by building ever more sophisticated equipment,
know the position and velocity of every star in the Milky Way.  Stars in tidal streams are 
the only stars which one can even in principle know about their positions and velocities in
the past.  Stars in tidal streams all resided, at one time in the past, in the progenitor
satellite dwarf galaxy or star cluster.  It is this property of tidal streams that allows
us to use them to constrain the Milky Way potential.

Because the Sgr dwarf tidal tails are the most prominent tidal debris in the spheroid, and
because the Sgr dwarf orbital plane is approximately perpendicular to the Galactic plane,
observations of the Sgr tidal debris potentially provide a strong indication of the shape of the gravitational
potential surrounding the Milky Way.  Interestingly, a comparison of disruption models with stellar
positions and velocities in the Sgr tidal tails has produced a contradictory result.
The leading and trailing Sgr tidal tails do not lie on the 
same plane \citep{nyetal03}; they are tilted in opposite directions with respect to the
orbital plane of the Sgr dwarf galaxy.  Their tilt suggests that the Galactic potential is 
oblate \citep{mgac04,jlm05,ljm05}.  However, the radial velocities of 
what appear to be some leading tidal tail stars 
can only be fit to disruption models with prolate Galactic potentials \citep{h04,ljm05}.  
The leading tidal tail stars that are not compatible with oblate Galactic potentials
are near the various identified substructures in Virgo.

One must consider the possibility that the Milky Way potential is more complex than the models, or 
that some Sgr leading tidal tail stars might be misidentified.  \citet{metal06} seems to suggest 
the latter.  Figure 3 of their paper shows radial velocities for leading tidal tail debris 
that has $V_{gsr}$ measurements like those of an oblate dark matter halo, and stars with 
$V_{gsr}\approx -75$ km s$^{-1}$ that they interpret as arising from an ancient episode of 
tidal stripping.  \citet{ljm05}, however, show the leading tidal stream from Sgr
transitioning seamlessly into a clump of stars with $V_{gsr}\approx -100$ km s$^{-1}$.  It 
is this latter clump of Sgr leading tidal tail stars that can only be fit with a prolate halo 
model.  The data in both papers come from 2MASS selected M giants, and the velocity dispersions of 
the stars in the vicinity of the Virgo constellation, with 
$V_{gsr}\approx -100$ km s$^{-1}$, are much larger than other places along the tidal debris 
stream.

Another complication that potentially blurs our understanding of the nature and extent of 
the overdensity (or overdensities) in Virgo is the suggestion that the smooth population of stars in the 
spheroid might be triaxial \citep{ny05,ny06,snfg06}.  A triaxial spheroid can produce a stellar 
density that is not symmetric around the center of the Galaxy, as viewed from our vantage point
near the Sun.  This idea was proposed because 
several authors see more spheroid stars in quadrant IV ($270^\circ<l<360^\circ$) than we
see in quadrant I ($0^\circ<l<90^\circ$), and because the lowest density of spheroid stars 
appears to be at a Galactic longitude slightly less than $180^\circ$ \citep{ny06}.  A triaxial spheroid
might explain how an overdensity could cover a large region of the sky and not immediately 
disperse; the overdensity in Virgo might be partially composed of an increased
density of spheroid stars from the smooth component.  Figure 3 of \citet{ny06} shows that
no model has been found that completely accounts for the observed overdensity of stars near
S297+63-20.5, so
although a triaxial spheroid might fit the tails of an overdensity in Virgo, we would still
expect to find a more spatially concentrated component of tidal debris with coherent velocities.

A key distinction between explaining an asymmetry with a `local' stream vs. a `global' structure
such as a triaxial spheroid is the difference in observed velocity dispersion of member stars.
Stars in a stream must have a relatively small velocity dispersion ($\sigma < 30 \rm \> km\>s^{-1}$),
 given the origin of the stream and given the fact that stream remains coherent as it moves
through the Galaxy's potential.   On the other hand, a triaxial spheroid structure in the halo
may have a very large velocity dispersion ($\sigma \sim 100 \rm \> km\> s^{-1}$), but
evidence for its member stars will be spread around the whole of the halo.

In this paper we attempt to disentangle the components of the spheroid substructure in 
Virgo using SDSS photometry of Blue Horizontal Branch (BHB) and F turnoff stars to build 
density maps of the tidal streams and the stellar spheroid itself.  We show that the Sgr
dwarf leading tidal tail, which is in the North Galactic Cap, arcs over (farther from the
Galactic plane) the S297+63-20.5 overdensity, and over the position of the Sun in the
Galactic plane.  If one extrapolates the path of the Sgr leading tidal tail below $b=30^\circ$,
it passes through the Galactic plane 15 kpc or more outside the Solar Circle.  This calls into
question both the identification of S297+63-20.5 stars with the Sgr leading tidal tail and 
the identity of the M giant stars identified with the leading tidal tail of Sagittarius
that have velocities in conflict with oblate models of the Galactic potential.

We show that there is a clear peak in the density of $20<g_0<21$ F turnoff stars near
$(l,b)=(300^\circ,60^\circ)$ that is at least $20^\circ$ across, but the total extent on
the sky and distance range cannot be constrained by photometry alone.
The densities of F turnoff stars over a four magnitude range ($18.5 < g_0 < 22.5$) are not consistent with 
an axisymmetric spheroid.  If the absolute magnitude range of the F turnoff
stars is 1.5 magnitudes, the spread in distance is a factor of 3, or in this
case 10-32 kpc.  In every direction that we probe at these implied distances, there are more stars in
quadrants III \& IV than there are in the symmetric position (with the same Galactic latitude
but with longitude $l_{symmetric} = 360^\circ-l$) in quadrants II \& I.  If the spheroid is axisymmetric,
then the number counts should differ only where there is a significant tidal substructure.  Either 
there is an enormous substructure centered in Virgo but covering a large fraction of the 
Galaxy or the smooth portion of the spheroid is not axisymmetric.

We then measure the velocities of the F turnoff stars associated with S297+63-20.5 and find
that their line-of-sight, Galactic center of rest velocity is $V_{gsr}=130\pm 10\rm \>km\>s^{-1}$.  
Throughout this paper, the symbol $V_{gsr}$ will refer to the line-of-sight component of the
velocity, as determined from radial velocity, transformed to the Galactic standard of rest,
using a Solar motion of $(v_X, v_Y, v_Z) = (10, 225, 7)$ km s$^{-1}$ \citep{db98}.
The measured $V_{gsr}$ is 
not consistent with the leading tidal tail of Sgr, which has negative $V_{gsr}$, in support
of our claim that S297+63-20.5 is not spatially coincident with the Sgr stream.  This velocity
is significantly higher than the measured $V_{gsr}$ for the VSS, but not so different that
the two structures could not be related in some way.  

In the direction $(l,b)=(300^\circ,55^\circ)$,
we see evidence for peaks with $V_{gsr}=-76\pm10\rm \>km\>s^{-1}$ and 
$V_{gsr}=-168\pm10\rm \>km\>s^{-1}$.  The first of the
two peaks appears to have a bluer turnoff than the latter peak, and could potentially be related to
the 2MASS M giant stars that have previously been assigned to the Sgr leading tidal tail.

\section{Spheroid Substructure from SDSS Photometry}

\cite{ynetal00}, \citet{nyetal02}, and \cite{nyetal03} show that both the leading and 
trailing tails of the Sgr dwarf spheroidal galaxy can be traced in SDSS color-selected BHB and 
F turnoff stars.  Since this work, much more SDSS data have become available, covering nearly 
the entire North Galactic Cap. We will use the BHB and F turnoff stars in these data to 
trace the Sagittarius dwarf galaxy tidal tails, and explore their connection with the S297+63-20.5
overdensity in Virgo.  The data presented in this paper come from the SDSS DR5 \citep{yetal00,dr5} plus about 
200 square degrees of imaging data from SDSS II that fills in the gap in DR5 photometry 
near the Galactic pole (to be released as part of SDSS-II DR6).  We will be analyzing 
$ugriz$ photometry \citep{figdss96, setal02} of 8500 sq. degrees in the North Galactic 
Cap. Details of the SDSS survey geometry may be found in \cite{setal01} and 
\cite{aetal03}.  Other SDSS technical information can be found in: \citet{getal98}, 
\citet{hfsg01}, \citet{pmhhkli03}, \citet{ietal04}, \citet{getal06}, and \citet{tetal06}.

\subsection{BHB stars - Tracing the Sgr Tidal Tails}

Blue Horizontal Branch (BHB) stars are important for tracing spheroid structure because
they are intrinsically bright and because they are all approximately the same absolute magnitude
so their distances can be estimated from apparent magnitudes to create a three dimensional
picture of the spheroid.  BHB stars are typically found in older stellar populations such as
are found in the spheroid, but are not found in all old populations.  Even in the spheroid,
there are a significant number of high surface gravity stars with colors that suggest a
spectral type of A.  These stars are either blue stragglers or young stars that have been
stripped from a star-forming region such as a dwarf galaxy, and are approximately two magnitudes
fainter than BHB stars.  From SDSS photometry we can select stars with colors consistent 
with spectral type A, and then we further divide this sample into
the stars whose photometry is more consistent with high surface gravity and the stars
whose photometry is more consistent with a BHB star.  The photometric separation of A stars
by surface gravity is not perfect, but is good enough to reveal important spheroid substructures.

Photometrically selected A-colored stars were chosen from the SDSS database in
a similar manner to \citet{ynetal00}.  The stars were selected by the 
criteria: ``PRIMARY," which eliminates duplicates, classified 
as ``STAR" because the point-spread 
function was consistent with a point source, and 
having colors $-0.3 < (g-r)_0 < 0.0$, and $0.8 < (u-g)_0 < 1.5$.  The 
subscript `0' indicates that the magnitudes were corrected by the 
\cite{sfd98} reddening as implemented for SDSS Data Release 5 (DR5; Adelman et al. 2007).  
This correction is appropriate given the large distance of the BHB stars.  We further 
separate higher surface gravity blue straggler (BS) stars from the 
lower surface gravity blue horizontal branch stars using the color 
selection criteria of Figure 10 in \citet{ynetal00}.

Fig.~\ref{sagxy} shows the spatial positions of the BHB stars in the plane of the Sgr
dwarf orbit, under the assumption that all of the stars have absolute magnitude 
$M_{g_0} =0.7$ \citep{ynetal00}.  We selected only BHB stars whose
projected position was within 15 kpc of the Sgr dwarf orbital plane, as defined in \citet{mswo03},
and rejected all stars within $0.2^\circ$ of the globular clusters M53, NGC 5053, NGC 4147,
and NGC 5466. The $X_{SGR,GC}$ and $Y_{SGR,GC}$ coordinates are in the Sgr orbital plane, 
and the Galactic plane is at $Y_{SGR,GC}=0$.  One can faintly see the newly discovered 
Bootes dwarf galaxy (Belokurov et al. 2006b) near $(X_{SGR,GC}, Y_{SGR,GC}) = (10,-55)$ kpc.
This figure can be compared with Figure 11 of \citet{mswo03}, which is reproduced in 
Fig.~\ref{f11} of this paper, and shows the positions of 
2MASS selected M giant stars in the Sgr tidal tails.  

In Fig.~\ref{sagxy}, one can clearly see the leading tidal tail of the Sgr dwarf tidal stream, 
30-40 kpc above the Galactic plane.  A piece of the trailing tidal tail is in the upper left 
corner.  The center of the leading tidal tail appears to arc over the
solar position, and to descend outside the solar circle if one extrapolates the
trajectory down toward the plane.  The stars in Sgr stream directly over the Sun in 
Fig.~\ref{sagxy} appear to be in a very broad region of the sky, which is not consistent with results of 
models of Sgr tidal debris, for example those of \citet{ljm05}.  Since we have not 
surveyed the entire volume within 15 kpc of the Sgr dwarf orbital plane, there are non-trivial 
selection effects near the edges of the data, which in particular make it difficult to study 
the structure within 20 kpc of the Sun in this figure.  The leading tidal tail appears unexpectedly
broad in this projection as it arcs over the Sun, and the extrapolation of the leading
tidal tail toward the Galactic plane is somewhat ambiguous.  

Since viewing the data in the observed (magnitude) rather the derived (distance)
space often yields higher contrast on any structures present in the halo, we show in 
Figure \ref{glambda} the $g_0$ magnitude vs. $\Lambda_\odot$ angular position along the 
Sgr tidal stream, separately for BHB and BS stars within 15 kpc of the Sgr dwarf orbital plane.
The left panel contains the same stars as Fig.~\ref{sagxy}, but in apparent magnitude
versus angle along the Sgr orbit.  The angle $\Lambda_\odot$ is measured from the center
of the Sgr dwarf galaxy, as viewed from the Sun, back along the direction of the trailing tidal
tail, as defined by \cite{mswo03}.  This is the longitude part of a $(\Lambda_\odot,B_\odot)$
system in which the equator is rotated to line up with the Sgr dwarf orbital plane.

A comparison of the left and right panels shows that the BHB/BS separation is quite 
good but not perfect.  Very few of the BHB stars in the leading tidal tail of Sgr are present
in the right panel showing high surface gravity (blue straggler) stars.  We 
have not analyzed the concentration of stars brighter than $g_0=15.5$.  
These stars are saturated in the SDSS images, and more care is required to 
determine how well color separation works with the interpolated 
magnitudes of these stars.  This figure can be compared with 
the 2MASS M giant stars in the Sgr tidal tails of Figure 8 
in \cite{mswo03} and Figure 2 of \citet{nyetal03}.  The $\Lambda_\odot$ axis is flipped from 
previous papers, but matches the orientation of the stream in Figure \ref{sagxy}.

There is a density peak in the left panel of Figure~\ref{glambda}, centered at 
$(\Lambda_\odot, g_0) = (240^\circ, 16.7)$, which we initially thought was related
to at least one of the previously discovered overdensities in
Virgo.  However, Figure \ref{glambda2} shows that this overdensity
is on the wrong side of the Galaxy for it to be in Virgo.
Figure \ref{glambda2} splits the data from the left panel of Figure \ref{glambda} into two pieces:
the left panel of Figure \ref{glambda2} shows stars with $Z_{\rm SGR,GC} < 0$ and to the right
shows that with $Z_{\rm SGR,GC} > 0$.  Since $Z_{\rm SGR,GC}$ is in approximately the same direction 
as the solar motion (Galactic $Y$), the left panel shows primarily stars in quadrants 
III and IV, and the right panel shows primarily stars in quadrants I and II.
The overdensity at $(\Lambda_\odot, g_0) = (240^\circ, 16.7)$ is in the right panel in Fig. \ref{glambda2}, which contains
primarily stars with $0^\circ<l<180^\circ$.  The origin of this overdensity is
still uncertain (but see Belokurov et al. 2007b).  It could be a single debris structure or a 
superposition of separate streams.  It is notable that several known globular clusters 
conspire to be in approximately the same position in this diagram; possibly there is 
debris associated with these globular clusters that is as yet unidentified.

Note from Figures \ref{glambda} and \ref{glambda2} that  the trailing tidal tail shifts from positive
$Z_{SGR,GC}$ to negative $Z_{SGR,GC}$ with increasing $\Lambda_\odot$.  In the left panel of 
Figure \ref{glambda2}, one notes a low contrast faint extension of this tidal tail down to 
$(\Lambda_\odot, g_0) = (250^\circ, 17.3)$.  In the right panel of Figure \ref{glambda2}
one sees the leading tidal tail at positive $Z_{SGR,GC}$.  This also decreases substantially in density
as one proceeds along it to the left, and appears to be close to $Z_{SGR,GC}=0$ so it is split between
the left and right panels in Figure \ref{glambda2}.  In both the leading and trailing tidal tails, we 
see the pile-up
of debris near apogalacticon.  In this projection, since the Sagittarius dwarf galaxy is $14^\circ$
below the Galactic plane, material falling down on us from close to the North Galactic Pole 
in this orbital plane would fall toward us at $\Lambda_\odot = 270^\circ - 14^\circ = 256^\circ$.  
Since the leading tidal tail arcs past
this angle, it likely passes through the Galactic plane outside the solar circle.

\subsection{F turnoff Stars - Separating Sgr from S297+63-20.5}

Since there are many more F turnoff stars in any stellar population than there are BHB stars, these F stars
are in principle better tracers of substructure.
The disadvantages of using 
F turnoff stars is that they are 3 or 4 magnitudes
fainter than the horizontal branch, and have a broader spread in absolute magnitude (so they are less accurate distance indicators).  The Sagittarius dwarf galaxy and tidal stream have a particularly blue
turnoff which makes them easy to separate from other Galactic populations.  In this section we present data which follows
the leading tidal tail from the Sagittarius dwarf galaxy and shows that it misses the position of
the S297+63-20.5 overdensity and also passes over the position of the Sun before
falling down on the
Galactic plane outside the solar circle.  We then present polar plots of F star density in the North Galactic
Cap in the magnitude ranges $19<g_0<20$, $20<g_0<21$, and $21<g_0<22$.  The polar plots show that the
Sagittarius leading tidal tail is fainter and offset from the S297+63-20.5 overdensity, and also show
the extent of the S297+63-20.5 overdensity.  

Fig.~\ref{f11} shows our detections of Sagittarius dwarf tidal debris from F turnoff stars in SDSS
stripes 13 and 15-23,
along with the previous detections of Sgr debris from BHB stars (Figure 3 of Newberg et al. 2003)
and 2MASS M giants \citep{mswo03}.
The new Sgr debris positions were determined by selecting by eye the center of the highest density peak 
in a histogram of $g_0$ versus angle along the SDSS stripe for blue F stars ($0.0 < (g-r)_0 < 0.3$). 
These positions were then converted to $(X_{SGR,GC},Y_{SGR,GC},Z_{SGR,GC})$, assuming $M_{g_0}=4.2$.  
This absolute magnitude was determined in \citet{nyetal02} by comparison with the Sgr BHB stars, so the SDSS BHB
star distances should be on the same scale as the SDSS F turnoff star distances, though there could be
an overall scale error.  Note that the Sagittarius M giants have systematically smaller heliocentric 
distances than either the A stars or the F stars \citep{chouetal06}.  

The debris was identified as Sgr tidal debris because it was contiguous with the overdensity in
the adjacent stripe.  There was some freedom in choosing the stream positions, as the profile 
of the overdensity was often asymmetric or bifurcated (Belokurov et al. 2006a).  The larger open 
squares in Figure~\ref{f11} show the positions of the higher density tidal debris.  If there was a 
second piece of tidal debris near the main Sgr leading tidal
tail in any stripe, it is shown in Figure~\ref{f11} as a smaller open square.  The positions of the 
Sgr F turnoff star detections are presented in Table 1.

Fig.~\ref{f11} clearly demonstrates that the leading tidal tail of the Sagittarius dwarf will 
pierce the Galactic plane outside the solar circle.  This is in contrast to the 2MASS M star observations
shown on the same figure.  
It is known that the stellar population in
the tidal tails is more metal poor and older than the stars in the core of the Sgr dwarf galaxy \citep{betal06}, so
it is reasonable to expect that the stellar populations would differ as a function of distance along
the stream.  However, one doesn't expect that stellar population differences would produce this much of a distance
difference between estimates from F turnoff stars and M giants.  Another possibility that should be
considered is that the M giants near the end of the leading tidal tail are not all related to Sgr.

Fig. \ref{f11} also shows that although S297+63-20.5,
indicated with a large cross, is close to the plane of the Sagittarius dwarf orbit, S297+63-20.5 is much 
closer to the Sun than the Sgr leading tidal debris.  
Any debris that is in the Sgr dwarf orbit is plausibly connected to this dwarf galaxy, but 
this significant overdensity in Virgo is  not an extension of the Sgr leading tidal tail, and there is
no obvious connection with Sgr.

The distance discrepancy between S297+63-20.5 and the Sgr leading tidal tail 
is demonstrated in another way by looking at polar plots of F star densities in
the North Galactic Cap at different apparent magnitudes.  We selected the stellar data from SDSS DR6
with $0.2<(g-r)_0<0.3$ and $(u-g)_0>0.4$.  The $(g-r)_0$ range was selected to be centered on the
turnoff color of S297+63-20.5, which is measured as 0.26 in \citet{nyetal02}.  
By excluding stars
stars redder than $(g-r)_0=0.3$, one avoids thick disk turnoff stars, even if they exist at
these faint magnitudes.  Fig.~\ref{polarplot} (lower panel) shows F stars with $20<g_0<21$; this magnitude 
range brackets the estimated apparent magnitude of the turnoff of S297+63-20.5.
The figure includes photometric data from SDSS, as well as some extra photometric data obtained
as part of the Sloan Extension for Galactic Understanding and Evolution (SEGUE), which consists of
$2.5^\circ$-wide scans at constant Galactic longitude which appear as radial extensions to our
sky coverage of the North Galactic Cap in this figure.

The stars in Fig.~\ref{polarplot} trace structure in 
the spheroid of the Milky Way at distances $\sim 14.5 - 23$ kpc from the Sun.  A larger density 
of F stars is represented by a darker shading in the equal-area polar plot, centered on the 
North Galactic cap. The Sagittarius leading arm is clearly visible running from 
$(l,b) = (205^\circ,25^\circ)$ to $(305^\circ,65^\circ)$.  More prominent is the S297+63-20.5 overdensity,
centered near $l=300^\circ$, and hugging the lower edge of the data.  Note that although there could be some
overlap in the spatial positions of some of the stars, the denser portions of the Sgr stream are separate from S297+63-20.5.
While most of the stars in
the S297+63-20.5 are contained within an area of the sky 15$^\circ$ in diameter, the tails of the distribution
could cover half of the North Galactic Cap.

In order to show conclusively that the polar plot cuts through S297+63-20.5, we perform the mathematical
equivalent of folding the data from the top half of the lower panel in Fig.~\ref{polarplot} down over the
line described by $l=0^\circ$, $l=180^\circ$, and subtracting it from the data on the bottom half.  In 
other words, data between $l = 0^\circ$ and $l = 180^\circ$ is subtracted from
the symmetrical point with the identical $b$ at $l' = 360^\circ-l$ and the residual is binned and plotted.   
The result, displayed in the top panel in Fig.~\ref{polarplot}, shows a localized structure near
$(l,b)=(300^\circ,64^\circ)$, which is at the same position as S297+63-20.5, within the errors of each
measurement.  This folding of the data should leave zero residual (except where there are streams or other halo substructure) if the spheroid is axisymmetric.  If the spheroid is
asymmetric, then more stars should have been subtracted over a wide area of
the upper panel, but it is difficult to make a smooth spheroid density function
that would include a strong peak like the one we see at $l=300^\circ$.

To clarify the relationship between S297+63-20.5 and Sgr, we show in Fig.~\ref{polarplot2} the F 
star polar plots for stars selected with $19<g_0<20$ and $21<g_0<22$.  The upper panel shows the 
brighter stars.
There is still a clear excess of stars at $l>180^\circ$ compared to the position in the sky that is symmetric
with respect to the Galactic center.  For example, there are many more stars at $(l,b)=(285^\circ,60^\circ)$
than there are at $(l,b)=(75^\circ,60^\circ)$.  However, there is not a clear peak in the star counts.
The asymmetry in star counts could either be an asymmetry in a smooth distribution of
stars or a dispersed stream over a large area of sky.  

The lower panel is dominated by a more distant piece
of the Sagittarius dwarf leading tidal tail, crossed at $b\sim 50^\circ$ with the much fainter Orphan
Stream \citep{betal07a}.  There are excess stars near $(l,b)=(300^\circ,60^\circ)$, but they are confused
with and overwhelmed by stars from the Sgr tidal tails at this magnitude.  A connection
between the Sgr tidal tail and S297+63-20.5 cannot be completely ruled out, but the bulk of the turnoff stars in the 
Sgr stream are at least a magnitude fainter ($g_0 \sim 21.5$) than the bulk of the 
turnoff of S297+63-20.5 ($g_0 \sim 20.5$).  
Models of Sgr debris such as that of \citet{ljm05} show some debris
stars near the position of S297+63-20.5, but not a strong peak of debris.  Either the models do not
adequately match the stellar debris, or S297+63-20.5 was not stripped from Sgr.

The polar plots show in a compelling way that the Sgr leading tidal tail intersects neither the
Sun nor the S297+63-20.5 overdensity in Virgo.  The turnoff stars in the Sgr dwarf tidal stream are 
fainter than $g_0 = 21.0$ near $(l,b)=(297^\circ,63^\circ)$ - they show up only in the faintest polar
plot on the lower panel of Fig.~\ref{polarplot2}.  The Sgr leading tidal tail is also
separated from S297+63-20.5 by about ten degrees in the sky (with significant overlap).  
As the leading tidal tail descends closer to the Sun we see the stream moving
toward the anticenter (Fig.~\ref{polarplot}).  In other words, it passes over the Sun in the North
Galactic Cap.  Close to the anticenter, the Sgr stream is fainter than $g_0=21$ 
(compare Fig.~\ref{polarplot} with Fig.~\ref{polarplot2} near the anticenter).
Near $(l, b)=(205^\circ, 30^\circ)$ the Sgr stream is visible
in all of the F turnoff star polar plots, but the star density is the greatest at this
position for $20<(g-r)_0<21$.  Since the the Str stream turnoff at $l=205^\circ$ is at $g_0=20.5$, its 
implied distance is 18 kpc from the Sun.  The Sgr stream is therefore at least 15 kpc from the 
Solar position and well beyond the Solar Circle as it descends toward the Galactic plane.

\subsection{F Turnoff Stars - separating S297+63-20.5 from the Smooth Spheroid}

We have already remarked that S297+63-20.5 seems to extend over a large range of magnitudes and
a significant fraction of the sky.  We now study whether that large extent could be partially
explained by an asymmetric (triaxial) smooth distribution of spheroid stars.

Fig.~\ref{fcountsplates} shows quantitative comparisons between the numbers of F turnoff stars as
a function of magnitude near S297+63-20.5 and those at a symmetric position in the Milky Way,
with the same Galactic latitude but on the opposite side of the Galactic center.
If the spheroid is axisymmetric, then one could find the number of stars in S297+63-20.5
from the difference between the star counts at $(l,b)=(297^\circ,63^\circ)$ and the star counts at
$(l,b)=(63^\circ,63^\circ)$.
In particular, Fig.~\ref{fcountsplates} shows a comparison of star counts at two places in Virgo: 
$(l,b)=(288^\circ,62^\circ)$ (upper panel) and $(l,b)=(300^\circ,55^\circ)$ (lower panel), with 
the corresponding position on the sky on the other side of the Galactic center.  

These two positions were selected because they match the positions of SEGUE spectroscopic fields in which
we will analyze radial velocity kinematics later in this paper.  Each sky area probed has a radius of 
$1.5^\circ$, which matches the footprint of a SEGUE spectroscopic plate.  In addition, F stars are 
selected with a color range: $0.2<(g-r)_0<0.4$ to match the color range with the highest
concentration of F stars in S297+63-20.5.
This redder color range will include a significant number of thick disk stars at magnitudes brighter than
$g_0=18$, but at fainter magnitudes there will be little contamination.  Since the 
thick disk (unlike the halo), is axisymmetric, there should be no effect on the difference in bright star counts 
due to thick disk stars between the Virgo selected regions and the symmetric reference regions.

Comparing the star counts at $(l,b)=(288^\circ,62^\circ)$, which is centered on the 
central knot of the VSS, with the star counts at $(l,b)=(72^\circ,62^\circ)$ in  the upper panel of Fig.~\ref{fcountsplates},
one sees that the star counts start to diverge at about $18^{\rm th}$ magnitude in $g_0$.  By $g_0=20.5$, there are
not quite a factor of two more stars in the VSS field than in the comparison field.  In the magnitude
range $19.4<g_0<20.0$, there are 560 stars in the $(l,b)=(288^\circ,62^\circ)$ field and only
357 stars in the corresponding reference field at $(l,b)=(72^\circ,62^\circ)$.  Of the stars in the
VSS field, 36$\pm$6\% are part of S297+63-20.5.  In the magnitude range $20.0<g_0<20.3$, there are
369 stars in the VSS field and 206 stars in the reference field.  Of the fainter stars in the VSS field,
44$\pm$7\% are part of S297+63-20.5.  This means that if
one selected a random turnoff star in the VSS field at $g_0=20.5$, there is nearly a 50\% chance of
it being a star from S297+63-20.5.  

The $(l,b)=(300^\circ,55^\circ)$ field in Virgo diverges from the reference field at 
$(l,b)=(60^\circ,55^\circ)$ at the even brighter magnitude of $g_0=17$ (lower panel of 
Fig.~\ref{fcountsplates}).  About a quarter of the stars with $17<g_0<19$ are part of the 
S297+63-20.5 excess in Virgo.  In the magnitude range $19.4<g_0<20.0$, there are 699
stars in the Virgo field, and only 440 in the reference field.  Of the stars in the Virgo field,
37$\pm$5\% are part of S297+63-20.5.  In the magnitude range $20.0<g_0<20.3$, there are 453 stars in the
Virgo field and 238 stars in the reference field.  Of the fainter stars in the Virgo field, 47$\pm$6\%
are part of S297+63-20.5.  At a magnitude of $g_0=20.5$, more than half of the stars in
the $(300^\circ,55^\circ)$ field should be part of the excess.  If the excess population has a unique
kinematic signature, as is expected for a stream or satellite, it should be quite prominent in any sample
of faint F turnoff stars.  We will return to these fractions when we examine the kinematic distribution
of selected F turnoff star spectra in these fields.

The wide range of magnitudes and sky coverage makes the overdensity in Virgo difficult to 
rationalize with our expectations for tidal
debris, even before we begin looking at the spectroscopy results.  We consider a non-axisymmetric
spheroid model as a possible alternative
explanation for the excess star counts in the Virgo region.  

To compare the observed star counts with those
predicted by the best currently available asymmetric model of the spheroid, the
Galactocentric triaxial Hernquist model of \citep{ny05,ny06,snfg06} is chosen.  In this model,
the center of the spheroid distribution is fixed at 8.5 kpc from the Sun with an assumed absolute
magnitude of blue F turnoff stars of $M_g=4.2$.  If the absolute magnitude is incorrect, then the overall
scale of all lengths and distances in the model will be off by the same factor.  See \citet{ny06} 
and \citet{snfg06} for parameter definitions and values for the triaxial Hernquist model.
The model was generated to fit SDSS F turnoff star counts with color $0.1<(g-r)_0<0.3$, which is a bit bluer 
than the data at $0.2 < (g-r)_0 < 0.4$. To account for this,
the normalization of the model is adjusted upward by a factor of 1.5 so that it matches the star counts as 
well as possible.  We cannot correct for the fact that the model does not include the thick disk stars
at brighter magnitudes, so we will have to ignore the difference between the star counts and the model
brighter than about $g_0=18$.   These triaxial spheroid model curves are plotted in Fig.~\ref{fcountsplates}
for comparison with the F turnoff star counts.

The model is a reasonable fit to the data for $18<g_0<21$.  At brighter magnitudes the thick disk dominates,
and at fainter magnitudes there are fewer stars than the model predicts.  The discrepancy at faint magnitudes
is probably due to a deficiency in the models, and possibly the lower completeness
of the fainter stellar data.
Note that if one compares the star counts at 
$(l,b)=(288^\circ,62^\circ)$ and $(l,b)=(300^\circ,55^\circ)$ with the model, the star counts begin to
diverge somewhere between $g_0=19$ and $g_0=20$.  There is a much smaller excess in the number
of stars at magnitudes
brighter than $g_0=20$.  Even using the current best fit triaxial spheroid model, however, one cannot account
for all of the excess stars, particularly those with magnitudes $20<g_0<21$.

In the next section the two possibilities for explaining the stellar count excess are
explored using additional information obtained from a sample of spectra from the same population of
stars.  If the spheroid is axisymmetric, and the S297+63-20.5 structure is tidal debris, then 
37$\pm$4\% of the stars with $19.4<g_0<20.0$ should have coherent velocities (i.e. show a small $\sigma < 30 \rm \> km\> s^{-1}$ dispersion), and 46$\pm$5\% of the stars with
$20.0<g_0<20.3$ should have coherent velocities.  If the spheriod is asymmetric, then the fraction of
stars with coherent velocities is expected to be lower.  With the best fit model presented, one estimates that at
$(l,b)=(288^\circ,62^\circ)$, 64 of 560, or 11\% of stars with $19.4<g_0<20.0$ should have coherent
velocities, and 159 of 369, or 43\% of stars with $20.0<g_0<20.4$ should have coherent velocities.  In
the direction $(l,b)=(300^\circ,55^\circ)$, 76 of 699, or 11\% of the stars with $19.4<g_0<20.0$ should 
be coherent, and 127 of 453, or 28\% of stars with $20.0<g_0<20.4$ should be coherent.  The model numbers
are an estimate, and will change somewhat as new non-axisymmetric spheroid models are developed.

\section{Substructure in Velocities of F Turnoff Stars}

Since S297+63-20.5 was discovered in F turnoff stars, and is prominent against the background of
smooth spheroid stars in that population, one would like to measure the velocities
of these same types of stars if possible.  Since the overdensity is most noticeable in quite 
faint F turnoff stars (with $20.0 < g_0 < 21.0$), this is not an easy task.  Essentially zero 
F turnoff stars in this magnitude range were observed in 
the primary SDSS galaxy survey.  

We are currently in the middle of SDSS II, which includes three survey projects, including the
Sloan Extension for Galactic Understanding and Exploration (SEGUE).
The SEGUE portion of SDSS II is designed to reveal
velocity substructure in the Milky Way thick disk and spheroid through analysis of radial
velocities and stellar properties from spectroscopy of $\sim 250,000$ Galactic stars.  
SEGUE spectra are identical
to SDSS spectra, in that they are observed with the same instrument and processed with the same (evolving)
software; the spectra cover 3700-9300 \AA\ with a resolution of $\approx$ 2000.  The difference is that SDSS
targeted a very few Galactic stars, most of which are BHBs or bright F subdwarfs that were used as standard
stars, while SEGUE targets Galactic stars exclusively.  While the SDSS obtained galaxy spectra in all areas
of the sky with imaging data, SEGUE samples the sky with a set of pencil beam spectral surveys 
in 200 target fields, spaced roughly 10 to 20 degrees apart over 3/4 of the sky.  

In the SEGUE survey, two 640 fiber plates of spectra \citep{setal01} are obtained in each target field.
The brighter targets are put on one plate which is exposed about 45 minutes.  Stars fainter than 
about $r_0 = 17.8$ are assigned to the second plate, which is exposed for 1.5 to 2 hours.  
The faint plates reach about $g_0 = 20.5$, which is just at the bright end of the S297+63-20.5 turnoff population.
Targets for SEGUE spectroscopic fibers are chosen by color from the SDSS imaging data to represent 
a variety of stellar types at a variety of distances sampling the thick disk and spheroid of the Milky Way.

We have already obtained two pairs 
of SEGUE spectroscopic plates in directions that probe the S297+63-20.5 overdensity.
These plate pairs are numbered 2689/2707 (bright/faint) at $(l,b)=(300^\circ,55^\circ)$ and 
2558/2568 at $(l,b)=(288^\circ,62^\circ)$.
The second plate pair is centered on the VSS RR Lyrae star overdensity.  
During DR6 spectro test processing of globular cluster calibration stars with known cataloged 
velocities, a universal offset of $7 \rm \>km\>s^{-1}$ was noted in the 
wavelength solutions; this zero point correction was added to all SEGUE radial velocities used 
in this paper.

Most of the faint stars of interest, with $0.1 < (g-r)_0 < 0.4$, are selected as ``F subdwarf,"
``Blue Horizontal Branch," or ``F/G" spectroscopic target types.  The BHB category goes as red 
as $(g-r)_0 < 0.2$ and as faint as
$g_0 < 20.5$.  These stars are randomly sampled to favor lower surface gravity stars using a photometric
measure of surface gravity for A-colored stars called the v-index \citep{lnrrs98} to estimate surface
gravity.  The F/G category randomly samples stars with $r_0<20.2$ and $0.2<(g-r)_0<0.48$.  The faint subdwarf
category favors F stars with low metalicity by favoring stars that are at the very bluest tip of the
stellar locus in $(u-g)_0$ and $(g-r)_0$, using a measure of position along the stellar locus in color space called $P1$
\citep{hetal03}.  The criteria for selecting each SEGUE category are complex, and include techniques for
varying the likelihood of selecting a star for spectroscopy as a function of apparent magnitude, so that
we select fewer of the more numerous, fainter stars.  The broader goal is to sample all of the components
of the Milky Way, so less common components are favored in the selection process.

Fig.~\ref{seguets} shows the color magnitude diagram of the bluer SDSS stars within 
$1.5^\circ$ of the center of the SEGUE plate pair 2689/2707, along with the colors and magnitudes
of the stars with $0.1<(g-r)_0<0.4$ and $19.4 < g_0 < 20.5$
for which we have obtained SEGUE spectra with the
two SEGUE plate pairs 2689/2707 and 2558/2568.  This is the entire region from which the spectra in
this paper are selected.  The symbols show that the spectra
are spread in color and magnitude, and do not follow the color distribution of the spheroid stars.
We must select stars in narrow color and magnitude ranges that correspond to a component we intend to
sample; we cannot rely on the fact that one component has many more stars to guarantee that it will
dominate the sample.

The stars that we expect to best sample the turnoff of S297+63-20.5 are those with $0.2<(g-r)_0<0.4$
and $20<g_0<21$.  The SEGUE spectra only go as faint as $g_0<20.5$, and most of the spectra have
$g_0<20.3$, so we sample only the brightest end of the magnitude range.  The right panels of
Fig.~\ref{fstarrv} show $V_{gsr}$ for stars from SEGUE faint plates 2707 and 2568 (available publicly with 
the release of SDSS DR7) that have the colors and magnitudes of S297+63-20.5 F turnoff stars, and
radial velocity errors, as measured by cross-correlation with ELODIE
standard templates \citep{mipc04}, of less than 20 km s$^{-1}$.  

In both plates, there is a clear excess of faint stars with positive radial velocities.  In the combined
panel (bottom right of Fig.~\ref{fstarrv}), there is a five sigma peak in the bin $112<V_{gsr}<144 \rm \>km\>s^{-1}$.  There
are 30 stars with positive $V_{gsr}$ and only 13 stars with negative $V_{gsr}$.  If all of the excess stars
are associated with S297+63-20.5, then 40$\pm$11\% of the fainter F turnoff stars could be associated
with a single tidal debris structure in Virgo.  

This fraction is consistent with estimates from photometry, whether or not the spheroid is symmetric.
In the symmetric spheroid case, we expected $46\pm5$\% of the excess stars with $20.0<g_0<20.3$ to be
in the excess population.  If the triaxial spheroid model is correct, the percentage in the excess
population is about 35\% for both plates combined.  Our measured value of $40\pm11$\% is consistent with
either of these options.  Therefore, from the faint ($g_0 > 20$) spectroscopic and photometric
data combined we conclude: the majority of the stars in the S297+63-20.5 overdensity are consistent with the presence of single substructure with a coherent velocity of
$130 \pm 10 \rm \> km\>s^{-1}$ and relatively small dispersion, as we would expect for tidal debris.  

In the left panels of Fig.~\ref{fstarrv}, we show $V_{gsr}$ for the stars with $19.4<g_0<20.0$.  In these
plots, there is no excess in the $122<V_{gsr}<144 \rm \>km\>s^{-1}$  bin.  One plate has more positive velocities, and the
other has more negative velocities.  In the combined panel (lower left of Fig.~\ref{fstarrv}), the only
significant peak has $-176<V_{gsr}<-144\rm \>km\>s^{-1}$.  Thus the brighter F stars, in the left panels of 
Fig.~\ref{fstarrv}, have distinctly different kinematics than the fainter stars in the right panels
of the same figure.  Out of 72 brighter stars, a few at best are part of the moving group we 
identified in fainter stars.  The symmetric spheroid assumption indicated that we should have found 37\%,
which is 27 stars, in a peak near $122<V_{gsr}<144\rm\>km\>s^{-1}$.  The most we can imagine assigning to a peak
are about seven stars from the plate at $(l,b)=(288^\circ,62^\circ)$ that are around $V_{gsr}=200\rm\>km\>s^{-1}$.
The moving group is ten percent
or less of the stars with $19.4<g_0<20.0$.
This differs at the $4\sigma$ level with our expectation of 27 stars present in a peak if the halo is axisymmetric.  

This measurement of brighter ($19.4 <g_0 < 20$) star kinematics and photometry
thus favors a triaxial spheroid.  In the symmetric spheroid model (plus individual star stream peaks), we expected $37\pm 4$\%
of the stars to be coherent in velocity, and we only found 11\% at the most.  The triaxial spheroid model
predicted that 11\% of the stars would be coherent in velocity, which is in good
agreement with the observed fraction.

There is one significant peak in the velocities of the brighter F turnoff stars, but it has a very
negative $V_{gsr}$.
We identify as a ``peak" any bin with more than a $2.5\sigma$ excess in the expected number of stars, 
where $\sigma^2=y(1-y/n)$, $y$ is the expected number of stars, and $n$ is the total number of stars 
in the 100 square degree region.  The limit for a bin to be identified as a peak is shown by the dotted
curve in Fig.\ref{fstarrv}.  In plate 2707, there are five stars with $-176<V_{gsr}<-144 \rm km\>s\>^{-1}$ where we would
have expected only one, and three stars in plate 2568 where we would have expected only two.  This amounts
to an excess of about five stars in the summed histogram, lower left of Fig.~\ref{fstarrv}.
The mean velocity of this peak is $-168\pm 10\rm \>km\>s^{-1}$.

Since \citet{nyetal02} measure a turnoff color for S297+63-20.5 of $(g-r)_0=0.26$, one would not expect
many stars in this structure to have measured colors of $0.1<(g-r)_0<0.2$ in the SDSS.  We show
measurements of $V_{gsr}$ for these bluer stars in Fig.~\ref{fstarrv2}.  In the fainter ($20.0<(g-r)_0<20.5$)
set of stars, there are maybe two excess stars in the bin one to the right of the moving group
associated with the center of S297+63-20.5, but no significant velocity peaks are observed.

In the left panels of Fig.~\ref{fstarrv2}, we see two marginally significant peaks: one at nearly the same
velocity as the $V_{gsr}=130 \rm \>km\>s^{-1}$ peak in Fig.~\ref{fstarrv2}, and one at in the bin with 
$-80<V_{gsr}<-48$.  We looked at the individual stars in positive velocity peak, and most of them are 
at the very faint and very red edges of the selection criteria.  Possibly, they represent blue 
stragglers that are at the same distance but have brighter absolute magnitudes than the F turnoff 
stars in the moving group identified in the fainter panels of Fig.~\ref{fstarrv}.

The peak with negative radial velocity is at $-76\pm10\rm \>km\>s^{-1}$, but note that there are more
stars on the negative velocity side of this peak than on the positive side.  If most of the brighter stars
in this color range are associated either with the positive or the negative $V_{gsr}$ peak, then
the Gaussian distribution from the expected spheroid would be much lower and the center of the
peak would move one bin to the left.  The center of the peak is nearly between two bins. 
This peak is more pronounced in plate 2707 than in plate 2568.
Once this peak is identified, one can see that it also present at some level in the fainter panels on
the right of Fig.~\ref{fstarrv2}.  
Since the stars in this peak are at about the same distance and line-of-sight velocity as stars
which other authors have attributed to the Sgr dwarf leading tidal tail, they will be analyzed at
greater length in the discussion section of this paper.

\section{Discussion}

\subsection{Is S297+63-20.5 Related to the Sagittarius Dwarf Tidal Stream?}

There was a time when any significant clump of tidal debris in the spheroid was suspected of belonging to
the one known example of present-day merging in our galaxy $-$ the Sagittarius dwarf tidal stream.  We now 
know there are many tidal debris streams, though the only currently known tidal stream as large as that of the Sagittarius dwarf 
is the Monoceros stream in the Galactic plane \citep{ynetal03,ietal03}.  We identify stars in the tidal tails of the Sagittarius dwarf spheroidal galaxy 
because either their locations and radial velocities are contiguous to those of other known pieces of Sagittarius 
debris or because their locations and velocities match those of models.

In this paper, we have presented a strong argument that S297+63-20.5 is not part of the Sgr leading tidal tail
because the turnoff stars in S297+63-20.5 are much brighter than those in the leading tidal tail 
debris in the same direction in the sky, and the stars in S297+63-20.5 have positive $V_{gsr}$ while Sgr leading
tidal debris would have negative $V_{gsr}$.  Moreover, \citet{nyetal02} showed that the turnoff of
S297+63-20.5 is redder than that of the Sgr tidal stream anywhere it has been detected in the sky.

Although \citet{mpjai06} were able to generate a model that shows the leading tail of the Sagittarius 
stream going through the position of the S297+63-20.5, they did so at the cost of not fitting some of the 
more distant spatial detections of that same tidal tail.  In this paper, (see \S 2), we locate additional 
pieces of what appears to be the leading tidal tail of Sagittarius in SDSS F turnoff stars, and show 
that it does not intersect the position of S297+63-20.5.

Since the Sgr tidal debris may be bifurcated, lumpy, wrap multiple times 
around the Milky Way, and have different stellar populations in different 
places, it is difficult to prove that any spheroid substructure, especially one that happens
to be in the Sgr orbital plane, is unrelated to this tidal disruption event.  Since the S297+63-20.5 debris is outgoing in velocity,
it is possibly related to the Sgr trailing tidal tail.  However, no existing model shows a lump of material near
S297+63-20.5.  More extensive modeling and mapping of the Sgr tidal debris is required 
to determine definitively whether they are related to each other.

\subsection{Relationship between S297+63-20.5 and other Virgo Substructures}

What is much murkier is the relationship between S297+63-20.5 and the other substructures that have
been discovered in the Virgo region, and have been identified as overdensities in Virgo.
These include the VSS (aka the ``12$^{\rm h}$.4 Clump"), and the VOD.

The VSS is in the same general area of the sky as S297+63-20.5, and at a similar calculated distance.  They both
have positive $V_{gsr}$, though the VSS has a measured $V_{gsr}$ of $99.8\rm \>km\>s^{-1}$ compared to our measurement
of $130\rm \>km\>s^{-1}$ for S297+63-20.5.  It is possible that membership or measurement errors of a few 
stars in either the S297+63-20.5 sample or the VSS sample could explain the difference in measured
velocity.  The central knot of the VSS is also a bit offset ($9^\circ$) from the center of 
S297+63-20.5.  One wonders whether the clump of RR Lyrae stars that defines the VSS could be a 
substructure within S297+63-20.5, such as a disrupted GC within a disrupted dwarf galaxy.  Or, we may
find as the number of known stars in the VSS grows, it might resemble the S297+63-20.5 structure
more closely.  Given the prevalence of velocity substructures in even the small amount of data analyzed
here, it cannot be completely ruled out that the VSS and S297+63-20.5 might be chance 
superpositions.  However, that possibility seems fairly unlikely since the velocities are quite similar.
We expect that as more data is analyzed, the S297+62-20.5 overdensity will become known as the Virgo
Stellar Stream (VSS).

It is somewhat difficult to assess the relationship between S297+63-20.5 and the VOD.  In the
\citet{jetal06} paper, their first step is to determine the distance to each star in the SDSS 
photometric database.  All
subsequent analyses are carried out with these spatial positions.  The VOD has a distance of $\sim 5-15$
kpc, which is a little closer than our estimate of 18 kpc for S297+63-20.5, which is at a similar
estimated distance as the VSS.  However, our distance errors to S297+63-20.5 are large, and there is
no guarantee it is the same as the VSS.  
In particular, there is no guarantee that our absolute magnitude estimate of
$M_g=4.2$, which was derived for Sgr tidal debris with similar colors, is 
appropriate for S297+63-20.5.
The factor of 2 density
excess for the VOD is similar to the density excess that we see in comparisons of the number of F
turnoff stars near $(l,b)=(300^\circ,60^\circ)$ and $(l,b)=(60^\circ,60^\circ)$.  Also supporting the
identification of S297+63-20.5 with the VOD is the position in the sky and the large sky area covered.

If there were two separate overdensities, one at 10 kpc and one at 18 kpc, we would expect to see
a separate VOD structure in the top panel of Fig.~\ref{polarplot2}, with an apparent magnitude
around $g_0=19.2$.  All we see in that figure is a general asymmetry in the star counts - no
detected peak density.  This suggests that either our technique of estimating distance gives
different results from \citet{jetal06} or that the turnoff of the VOD is redder than $(g-r)_0=0.3$.
Probably, they are the same overdensity measured with the same star catalog, but it is difficult to
know for sure.  What is needed is a representation of the VOD stars that allows us to identify which
stars are members of the structure, such as a color magnitude diagram and a map of angular position
on the sky.  Then we would be able to check whether we are looking at the same stars in the same 
catalog or not.

\subsection{The Sgr Leading Tidal Tail and the Shape of the Milky Way Gravitational Potential}

The controversy regarding the oblateness of the Milky Way gravitational potential hinges on the
radial velocity data for M giants in the part of the leading tidal tail that is closest to the Sun,
$240^\circ<\Lambda_\odot <  260^\circ$.  
Figure 12 from \citet{ljm05} shows the model distances and velocities of Sagittarius debris 
for prolate, spherical, and oblate dark matter halos, along with data for 2MASS M stars.
There are a large number of stars near $\Lambda_\odot = 250^\circ$ that have
velocities of $-150 < V_{gsr} < -50$ km s$^{-1}$ and distances from the Sun of about 15 kpc.  It is 
precisely these stars that do not fit the
published models, unless one assumes a very prolate dark matter halo \citep{h04}.  
The stars with $200^\circ < \Lambda_\odot < 275^\circ$ also have a larger velocity dispersion than
Sgr debris stars at other places in the stream.

The last column in Table 1
gives the position of the Sgr leading tidal tail in terms of $\Lambda_\odot$; at $\Lambda_\odot = 250^\circ$
the distance to the Sgr stream is 36 kpc.  M giant
distances in \citet{mswo03} are systematically $13\%$ less than ours, as determined from a comparison
of M giant stars and A/F stream position of the leading tidal tail stars near apogalacticon
\citep{nyetal03}.  Applying
this scale difference, the M giant stars at $\Lambda_\odot = 250^\circ$ should be at 31 kpc
on the M giant distance scale.  There are very few stars in the \citet{ljm05} sample that are that
far away.

Fig.~\ref{f11} shows where the discrepancy arises.  If one looks directly up from the Sun in
Fig.~\ref{f11}, following the direction of $-Y_{SGR, GC}$, the Sgr F turnoff stars are estimated to be
more than 30 kpc above the Galactic plane.  In contrast, the M giant stars are scattered from 10 kpc 
to 35 kpc in the same direction.  The M giant stars are at about the same distance, and in about the
same direction, as the S297+63-20.5 overdensity.

In fact, we detect a radial velocity peak for F turnoff stars near S297+63-20.5 that has a velocity
of $V_{gsr}=-76\rm \>km\>s^{-1}$, which is very similar to the M giant velocities.  There is an excess of stars
in the bin with $-80<V_{gsr}<-48\rm \>km\>s^{-1}$ in every histogram in Fig.~\ref{fstarrv2}.  There are 15 of these 
bluer F turnoff stars ($0.1<(g-r)_0<0.2$) with $19.4<g_0<20.5$ where we expected to see less than
seven.  The excess in this bin is more than 3 sigma, and there are additional extra blue F turnoff
stars with more negative $V_{gsr}$.

There is also velocity peak in RR Lyrae stars near the VSS, at about the same distance and direction.
Figure 2 of \citet{detal06} shows spectroscopy of 28 BHB and RR Lyrae stars from the QUEST survey.
In the \citet{detal06} paper, a very tight
group of stars at 18.5-20 kpc from the Sun and within $10^\circ$ of each other in the sky is identified.
The mean velocity in the Galactic rest frame of these stars is 99.8 km s$^{-1}$.  Additional 
stars as close as 16 kpc and as distant as 24 kpc have similar enough radial velocities that they might 
be considered to be part of the same kinematic structure.

The remaining stars are not consistent with a Gaussian distribution, and we note that there is a
very significant peak in the lower panel of their Figure 2 that is very near $V_{gsr}=-75\rm \>km\>s^{-1}$.
If one considers all of the radial velocities obtained by \citet{detal06},
there are 12 stars in a peak at nearly 100 km s$^{-1}$, nine stars in a peak at $\approx -80$ km s$^{-1}$, 
and 7 stars with very large negative or positive radial velocities that are consistent with neither peak, 
nor with any existing model of spheroid kinematics.  There is insufficient information in the 
\citet{detal06} paper to adequately analyze the nine stars with velocities toward
the Sun, but they likely cover an area $20^\circ$ on the sky or larger, and at least the distance range
$14 < r_{\rm Sun} < 20$  kpc.  These stars are very likely associated with our F turnoff star peak
at $V_{gsr}=-76\rm \>km\>s^{-1}$ and with the stars that \citet{ljm05} assigns to the Sgr leading tidal tail.

Figure 3 of \citet{metal06} also show stars in the same part of the sky, $\Lambda_\odot = 239^\circ$,
and with similar velocities, $V_{gsr} \sim -100$ km s$^{-1}$. Interestingly, they identify these stars 
as part of an ancient tidal debris stream that was stripped from Sgr $>2$ Gyr ago.  Although 
this presents a possible explanation, it does not explain why the dispersion of the stars appears 
so large in this portion of the sky, and is at odds with the general assumption that 2MASS giants 
favor younger, more recently stripped portions of the debris.  Figure 12 of \citet{ljm05} shows a 
subset of stars in this region that have a very narrow velocity dispersion and very nearly match the 
expected position of the ancient tidal debris if one assumes an oblate dark matter halo.  One
wonders whether a simpler explanation for the debris with $V_{gsr}\approx -100$ km s$^{-1}$ might be 
the presence of a previously unidentified spheroid component.

The fact that the stars in this infalling structure have a blue turnoff supports, but does not prove,
the claim that they are part of the Sgr dwarf tidal stream.  However, they appear to be closer than
the main part of the tidal tail.  Further investigation is required to determine whether they are part
of the Sgr dwarf tidal debris, and whether they represent the population of stars they are being
compared to in the tidal disruption models.

\subsection{The Shape of the Spheroid}

If we knew the spatial position of every spheroid star, or at least a representative sample of spheroid
stars, we could use this information to construct a density model for this Milky Way component.
But given that there are very large known overdensities in the spheroid, what would the parameters in
this model be?  \citet{betal07} show that all smooth models are poor fits to the distribution of stars
in the spheroid.

We have chosen to model the spheroid as a set of significant overdensities that are coherent in
position and velocity, plus a roughly smooth spheroid distribution.  Overdensities include globular
clusters, dwarf galaxies, and large chunks of associated debris from these objects.  Thus, we are now
in the process of identifying large, coherent debris structures so that we can fit the remaining stars
to a smooth spheroid model; and fitting smooth spheroid models so we can identify overdensities that
might be large, coherent debris structures.

There is no controversy over the asymmetry of the star counts in the stellar spheroid stars in the
North Galactic Cap \citep{ny05,ny06,xdh06}, though there is a question of whether there is 
an asymmetry in the south \citep{xdh07}.  An important question is whether the asymmetry is due to
a single large debris structure added to an otherwise axisymmetric distribution of stars.

The data in this paper suggest that is not the case.  The S297+63-20.5 is a very significant, coherent
overdensity that contains about half the spheroid stars at $g_0=20.5$ and $(l,b)=(297^\circ,63^\circ)$.  However, 
we have shown that it is localized in magnitude and velocity, and does not explain the asymmetry at brighter magnitudes.

It is possible that there is more than one structure, or maybe a set of structures, that conspire to
make the North Galactic Cap appear asymmetric.  It seems at least as likely that the spheroid itself
is not symmetric about an axis through the center of the Milky Way and perpendicular to the Galactic plane.
We have shown that our existing triaxial spheroid model is not perfect, but goes some way toward
predicting the number of spheroid stars in the observations.  We hope the model will improve as 
more large, coherent spheroid substructures are identified and as our fitting procedures evolve.

\section{Conclusions}

The Sgr leading tidal tail is traced through the North Galactic Cap
over the Center of the Milky Way, over S297+63-20.5, over the Sun, and at $b=30^\circ$ it is 
heading toward the Galactic plane at a Galactic longitude of $l=205^\circ$.  We expect that it 
pierces the Galactic plane well outside the solar circle.  We question whether some of the
M giant stars that have been compared to Sgr dwarf disruption models are at the right distance
to be members of the Sgr dwarf leading tidal tail.
The main part of the Sgr dwarf leading tidal tail is not spatially coincident with the
S297+63-20.5 overdensity in Virgo.

The stars in S297+63-20.5 have line-of-sight, Galactic standard of rest velocities 
$V_{gsr}=130\pm 10\rm\>km\>s^{-1}$.  The velocity dispersion
is difficult to estimate because we do not know for sure which stars are part of the structure, but
velocity dispersions of 10 $\rm km\>s^{-1}$ to 30 $\rm \>km\>s^{-1}$ are not unreasonable.
Since the Sgr dwarf tidal tail would have negative velocities in this part of the sky, we
determine that S297+63-20.5 is not associated with the leading tidal tail.  Because the color of the
turnoff, the density structure, and the $V_{gsr}$ velocities do not match any known or expected Sgr debris, we suggest that
it is instead part of a distinct merger event.

The Virgo Stellar Stream (VSS, Duffau et al. 2006) is at the same distance from the Sun as S297+63-20.5
and has a similar but not identical position in the sky and in $V_{gsr}$.  It could be a the same 
as S297+63-20.5 or a structure within S297+63-20.5, but the relationship is not certain.  

The Virgo Overdensity (VOD, Juri\'{c} et al. 2006) and S297+63-20.5 are in the same position in the 
sky and have a similar number of stars in excess of the smooth spheroid, but they are at a different 
distances from the Sun.  Probably, there is a scale difference between our distance measurements and
the VOD is the same as S297+63-20.5.

We show that the number densities of F turnoff stars are not symmetric about the Galactic center at
$g_0\sim 19.5$, and that this discrepancy is not due to S297+63-20.5.  Either the spheroid is asymmetric
about the Galactic center, or there are additional substructures that conspire to be on the same
side of the Galaxy as S297+63-20.5.

Finally, we note that in every figure showing the spatial and velocity distribution
in the spheroid, there are hints of extra unidentified substructure.  In particular, we note
an unexplained overdensity of BHB stars at $(\Lambda_\odot, g_0)=(240^\circ, 16.7)$,
and two additional moving groups near S297+63-20.5 with velocities $V_{gsr}= -168\pm10\rm \>km\>s^{-1}$ and $V_{gsr}=-76\pm10\rm \>km\>s^{-1}$.

\acknowledgments

HJN acknowledges funding from the National Science Foundation 
(AST-0307571, AST-0607618, AST-0612213), the NASA NY Space Grant, and John Huberty. 
TCB acknowledges partial funding for this work from National Science 
Foundation grants AST 04-06784 and PHY 02-16783: Physics Frontiers 
Center/Joint Institute for Nuclear Astrophysics (JINA).

Funding for the SDSS and SDSS-II has been provided by the Alfred P. Sloan Foundation, the Participating Institutions, the National Science Foundation, the U.S. Department of Energy, the National Aeronautics and Space Administration, the Japanese Monbukagakusho, the Max Planck Society, and the Higher Education Funding Council for England. The SDSS Web Site is http://www.sdss.org/.

The SDSS is managed by the Astrophysical Research Consortium for the Participating Institutions. The Participating Institutions are the American Museum of Natural History, Astrophysical Institute Potsdam, University of Basel, Cambridge University, Case Western Reserve University, University of Chicago, Drexel University, Fermilab, the Institute for Advanced Study, the Japan Participation Group, Johns Hopkins University, the Joint Institute for Nuclear Astrophysics, the Kavli Institute for Particle Astrophysics and Cosmology, the Korean Scientist Group, the Chinese Academy of Sciences (LAMOST), Los Alamos National Laboratory, the Max-Planck-Institute for Astronomy (MPIA), the Max-Planck-Institute for Astrophysics (MPA), New Mexico State University, Ohio State University, University of Pittsburgh, University of Portsmouth, Princeton University, the United States Naval Observatory, and the University of Washington.

\clearpage

\begin{deluxetable}{rrrrrrrr}
\tabletypesize{\scriptsize}
\tablecolumns{8}
\footnotesize
\tablecaption{SDSS Stripes with Sgr Stream F turnoffs}
\tablewidth{0pt}
\tablehead{
\colhead{Stripe} & \colhead{incl.\tablenotemark{a}} & \colhead{$l$} & \colhead{$b$} & \colhead{$g_0$} & \colhead{$d$\tablenotemark{b}} & \colhead{$\mu$\tablenotemark{c}} &\colhead{$\Lambda_{\odot}$\tablenotemark{d}}  \\
\colhead{Number} & \colhead{$^\circ$} &  \colhead{$^\circ$} & \colhead{$^\circ$} & \colhead{mag} & \colhead{kpc} & \colhead{$^\circ$} &\colhead{$^\circ$}}
\startdata
13 & 7.5 & 308.7 & 70.2 & 22.00 & 36 & 195 & 253\\
15 & 12.5 & 277.1 & 73.9 & 21.85 & 34 & 186 & 242\\
16 & 15.0 & 245.4 & 69.0 & 21.60 & 30 & 175 & 231\\
17 & 17.5 & 233.2 & 65.9 & 21.52 & 29 & 170 & 225\\
18 & 20.0 & 221.7 & 57.2 & 21.35 & 27 & 160 & 216\\
19 & 22.5 & 215.9 & 51.0 & 21.12 & 24 & 153 & 209\\
20 & 25.0 & 212.1 & 46.7 & 21.12 & 24 & 149 & 205\\
21 & 27.5 & 209.2 & 41.0 & 21.02 & 23 & 143 & 200\\
22 & 30.0 & 207.2 & 33.7 & 20.68 & 20 & 136 & 193\\
19\tablenotemark{e} & 22.5 & 229.1 & 75.6 & 21.75 & 32 & 179 & 232\\
20\tablenotemark{e} & 25.0 & 217.2 & 72.9 & 21.45 & 28 & 175 & 226\\
21\tablenotemark{e} & 27.5 & 209.0 & 67.5 & 21.25 & 26 & 170 & 220\\
22\tablenotemark{e} & 30.0 & 204.0 & 62.4 & 21.65 & 31 & 165 & 218\\
23\tablenotemark{e} & 32.5 & 201.8 & 53.2 & 21.40 & 28 & 156 & 209\\
\enddata
\tablenotetext{(a)}{Inclination of stripe relative to celestial equator, node is at $\alpha = 95^\circ$.}
\tablenotetext{(b)}{Inferred distance of stars from the Sun, assuming $M_g(\rm F) = +4.2$.}
\tablenotetext{(c)}{Angle along an SDSS ``Stripe" which navigates an inclined great circle on the sky with a pole at $(\alpha, \delta) = (95^\circ, 0^\circ)$.}
\tablenotetext{(d)}{Sagittarius plane azimuth}
\tablenotetext{(e)}{Indicates a secondary peak in the stripe, tracing a different density ridge.}
\end{deluxetable}

\clearpage
\figcaption {Spatial positions of BHB stars within 15 kpc of the Sgr dwarf orbital plane.  
Notice the arc of the leading tidal tail descending 
from 40 kpc above the Galactic plane and falling down on the solar position at $X_{Sgr, GC} = -8.5$ kpc. 
The overdensity near the Sun appears unexpectedly wide compared
to the width of the leading tidal tail at apogalacticon.
Part of the trailing tidal tail can be seen at $(X_{SGR,GC},Y_{SGR,GC})=(-80,-40)$ kpc.
The Galactic plane
is at $Y_{Sgr, GC} = 0$; the Sgr orbital plane is as measured by \citet{mswo03} and with the Galaxy-centered
${X, Y, Z}$ convention as described in \citet{nyetal03}.  
\label{sagxy}}

\figcaption {$g_0$ magnitude vs. $\Lambda_\odot$ for BHB (left) and BS (right) stars within 15 kpc of the
Sgr dwarf orbital plane.  
The A stars were separated into blue horizontal branch (BHB) and blue straggler
(BS) stars using the color separation defined by Fig. 10 of \citet{ynetal00}.
The Sgr leading and trailing tidal tails are marked as ``Leading" and ``Trailing," respectively. 
The leading tidal tail is evident in BHB stars from 
$(\Lambda_\odot, g_0) = (290^\circ, 19.0)$ and sloping down toward the center of the diagram.  If the leading tail came down on the Solar position from close
to the North Galactic pole, it would come down at $\Lambda_\odot=256^\circ,$ which is marked in the diagram.
Note that it instead passes this angle, heading towards the Galactic anticenter.  The Sgr leading tidal tail is also
evident in BS stars two magnitudes fainter; a small fraction of these BS stars are observed leaking into the BHB sample on the left panel.  The trailing 
tidal tail is evident from $(\Lambda_\odot, g_0) = (185^\circ, 20.3)$ and sloping down toward the center of 
the diagram.  There is a very large falloff in the number of BHB's in the trailing tail at $\Lambda_\odot=195^\circ$.
Stars brighter than about 15.5 magnitudes are saturated in the SDSS survey, and
therefore have less accurate magnitudes.  It is unclear whether the bright stars are BHB stars out to 10 kpc,
or BS stars closer than 4 kpc.  In the left panel, there is a density peak of unknown origin centered at $(\Lambda_\odot, g_0) = 
(240^\circ, 16.7)$.  BHB stars within $0.2^\circ$ of the four globular clusters M53, NGC 5053, NGC 4147,
and NGC 5466 were discarded before making this diagram, which reduced but did not remove this density peak.
\label{glambda}} 

\figcaption{Separation of the BHB stars on each side of the Sagittarius orbital plane.
We divide the BHB stars in the left panel of Fig.~\ref{glambda} along the plane of the Sagittarius dwarf orbit.
The plane of the orbit of the dwarf is close to the Galactic $(X,Z)$ plane, so most of the stars in the
left panel have $180^\circ<l<360^\circ$ and most of the stars in the right panel have $0^\circ<l<180^\circ$.
Stars from four globular clusters have been removed from the plot.
Note that the peak at $(\Lambda_\odot, g_0) = (240^\circ, 16.7)$ is not in the direction of S297+63-20.5, at
$(l, b) = (297^\circ, 63^\circ)$; it is on the wrong side of the Sgr orbital plane.
The $(\Lambda_\odot, g_0) = (240^\circ, 16.7)$ is
low enough signal-to-noise that when we further subdivided the data into bins that were 5 kpc wide in $Z_{SGR,GC}$
it could not be discerned.  Note also in this figure that the trailing Sgr stream is more evident in 
the right panel near apogalacticon (on the left edge) and then is more prominent in the left panel 
at brighter magnitudes.  The trailing tidal tail is faint, but possibly extends as bright as $17^{\rm th}$
magnitude at $\Lambda_\odot=256^\circ$.
\label {glambda2}}

\figcaption{Position of S297+63-20.5 compared to other detections of the Sagittarius tidal tails.  We 
reproduce Figures 3 and 4 from \citet{nyetal03}, which show the positions of the A-colored stars of the Sgr stream 
selected from eleven SDSS stripes (filled circles are leading tidal debris and open circles are trailing tidal debris),
and the positions of 2MASS M giants from Fig. 11 of \citet{mswo03} (one point for each star).  
The larger open squares show the positions of ten new detections of the leading Sgr tidal tail.  The smaller open
squares show the positions of other bits of tidal debris detected in the same SDSS stripes, which is
possibly due to superposition of young leading tidal debris and old trailing tidal debris
\citep{fetal06}.  The smaller squares
trace debris on the opposite side of the Sgr orbital plane.
The cross shows the position of S297+63-20.5, at 
$(l,b,R) = (297^\circ, 63^\circ, 18 {\rm ~kpc})$, where R is the distance from the Sun.  Note that although the center of S297+63-20.5
appears to be in the plane of the Sgr leading tidal tail, the distance does not match previous detections of the leading tidal tail.
\label{f11}}

\figcaption {F Star Polar Plot of the North Galactic Cap. 
The bottom panel shows stars with $0.2 < (g-r)_0 < 0.3$ and $(u-g)_0 > 0.4$ in the magnitude range
$20.0 < g_0 <21.0$.  The North Galactic Pole is in the center of the plot, and the outer circle is at $b=30^\circ$.
The distance between concentric circles of constant
Galactic latitude was stretched to preserve solid angle per pixel.  
Darker areas of the diagram contain more F turnoff stars.  All of 
the magnitudes were corrected using 
the reddening map of \citet{sfd98} before selection.  
The dark areas at low latitude and toward
the Galactic center represent the smooth portion of the Galactic spheroid.  The dark line from 
$(l,b) = (205^\circ, 25^\circ)$ to $(l, b) = (255^\circ, 70^\circ)$ is the leading tail of the Sgr tidal stream,
descending toward the Galactic plane as one moves to the left.  The Sgr stream turnoff is fainter than $g_0 = 21$
at the longitude of S297+63-20.5.  A pair of square brackets encloses the
primary area containing the S297+63-20.5 stellar excess.
The top panel shows a
subtraction of the pixels on the top half of the polar diagram, which does not contain any obvious tidal debris,
from the lower half, with the assumption that the star counts are symmetric about $l=0^\circ,180^\circ$.
One sees in the subtracted diagram that S297+63-20.5 peaks in this magnitude range near
$(l,b) = (300^\circ, 60^\circ$); the density decreases for $b < 64^\circ$ as one moves down the center 
``outrigger" (SEGUE) extension of photometric data at $l=300^\circ$.  The Orphan Stream \citep{betal07a} is visible
on the edge of the data at $l=255^\circ$.  The tidal tails of the Pal 5 globular cluster \citep{oetal03} are
near the edge of the data at $l=0^\circ$.
\label{polarplot}} 

\figcaption{Brighter and fainter F Star Polar Plots.  SDSS data is extracted as in 
Figure \ref{polarplot}, except we select closer (brighter, $19 < g_0 < 20$) F turnoff stars 
(upper panel) and more distant (fainter, $21 < g_0 < 22$) stars (lower panel).  The Monoceros structure 
is apparent as a density enhancement toward the Galactic anti-center ($165^\circ < l < 225^\circ$) in the
upper panel.  The Sagittarius stream and the S297+63-20.5 structures are
prominent in the lower panel, as is Palomar 5 at $l=0^\circ, b=45^\circ$.  The brighter (top) panel shows
stars approximately 9 to 14.5 kpc from the Sun.  We do not see evidence for a separate VOD at this
distance.  There is, however, a general asymmetry in the number of spheroid stars around the
Galactic center at $l=0$.  The fainter (bottom) panel shows stars approximately 23 to 36 kpc from the
Sun.  Since the Sgr stream at the anticenter is most prominent in Fig~\ref{polarplot}, it is
between 14.5 and 23 kpc from the Sun as it passes below a Galactic longitude of $b=30^\circ$.  
\label{polarplot2}}

\figcaption{Magnitude distribution of F Turnoff Stars in Virgo.  We show histograms of F star 
($0.2 < g-r < 0.4$) counts in mirror image
1.5$^\circ$ radius fields centered on:  $(l,b) = (288^\circ,62^\circ)$, upper panel, light line, the VSS field; $(l,b) = (72^\circ,62^\circ)$, heavy line: the mirror 
image Galactic field at $l' = 360-l$. In the lower panel: $(l,b) = (300^\circ,55^\circ)$, 
light line, the S297+63-20.5 field; and $(l,b) = (60^\circ,55^\circ)$, heavy line, its mirror image field.
The strong asymmetry in number counts between the quadrant IV fields over
their quadrant I mirror images is apparent at 
magnitudes $18.5 < g_0 < 22.5$ (8 kpc $< d < $ 45 kpc).   The matching 
curves indicate the predicted triaxial halo model counts associated
with each pointing.  Note that neither a triaxial model alone nor
a single halo stream alone can match all the strongly asymmetric data (see text).
\label{fcountsplates}}

\figcaption {Colors and magnitudes of SDSS F turnoff stars which have spectra in SEGUE.  The
smaller black dots show the colors and magnitudes of stellar objects in the part of
the S297+63-20.5 structure observed with SEGUE plate 2707.  The
turnoff of S297+63-20.5 is visible in the enhanced density of field stars
at $g-r \sim 0.26$ and $g_0 \sim 20.5$.  Larger 
symbols indicate stars with $0.1 < (g-r)_0 < 0.4$ and $19.4 < g_0 < 20.5$ for which spectra were obtained in the same
1.5 degree radius field targeted with plates 2568 and 2707.  The symbol type indicates the
SEGUE target selection category:  BHB, F sub-dwarf and F/G dwarf candidates are variously-sized filled circles;  squares indicate low 
metalicity candidates, and crosses indicate cool white dwarf candidates.
The targets selected for spectroscopy were heavily weighted toward
low metalicity (generally bluer) candidates, and do not
representatively sample the color distribution of all objects in
the field (black dots) when $g-r_0 < 0.2$.
\label{seguets}}

\figcaption{Galactic standard of rest velocities of F stars near the center of S297+63-20.5.
We selected from SEGUE plates 2707 and 2568 all 91 of the stars with 
$0.2 < (g-r)_0 < 0.4$ and $19.4 < g_0 <20.5$ that also
had radial velocity errors of less than 20 km s$^{-1}$.  In the upper panels, 
stars from plate 2707  $(l,b) = (300^\circ,55^\circ)$  are subdivided into
brighter ($19.4 < g_0 < 20$, upper left), and fainter ($g_0 > 20$) subsamples. 
The middle two panels show the same subsets for 
plate 2568 $(l,b) = (288^\circ,62^\circ)$. The lower panels sum the
two panels above them, for these two fields separated by $\sim 9^\circ$ degrees
 on the sky ($2-3$ kpc at the distances implied by these 
turnoff stars).   In each panel, the solid line histogram represents a 
Gaussian distribution of halo stars centered on $V_{gsr}=0$ with $\sigma=120$ km s$^{-1}$.
The Gaussians are normalized so that the area under the curve equals the number of
spectra in that panel minus the number of stars above the Gaussian in bins that
have more than a $2.5 \sigma$ excess.  The dotted line shows the limits for a $2.5 \sigma$
excess in each bin. Any data peak above the dotted line represents a peak of 
greater than 2.5 $\sigma $ significance.  There are two significant
peaks apparent in the lower panels, one at $v_{gsr} = -168 \> \rm km\> s^{-1}$ 
for closer, brighter stars in the lower left panel, and a very prominent
peak  at $v_{gsr} = 130 \>\rm km\> s^{-1}$ 
in the lower right panel (fainter stars) that is associated with the
S297+63-20.5 structure at an implied distance of $\sim 18$ kpc from the sun.
\label{fstarrv}}

\figcaption{Galactic standard of rest velocities for F stars near the center of S297+63-20.5.
115 stars were selected and displayed as in Fig.\ref{fstarrv}, except they are bluer,
$0.1<(g-r)_0<0.2$.  There are two significant peaks: one at $V_{gsr}\sim 150\rm \>km\>s^{-1}$ in the left
middle panel, and one at $V_{gsr} = -76 \> \rm km\> s^{-1}$  in the upper left panel.  They
are both present in the sum of these two panels at the lower left.
The outgoing (positive $V_{gsr}$) stars are possibly related to S297+63-20.5.  The origin of the incoming stars, which
are also present at some level in the lower right panel, is less clear, but these stars have
similar velocities to the stars which \citet{ljm05} fit to the Sgr leading tidal tail.  \label{fstarrv2}}

\clearpage

\setcounter{page}{1}

%\plotone{xybhb4n.eps}
\plotone{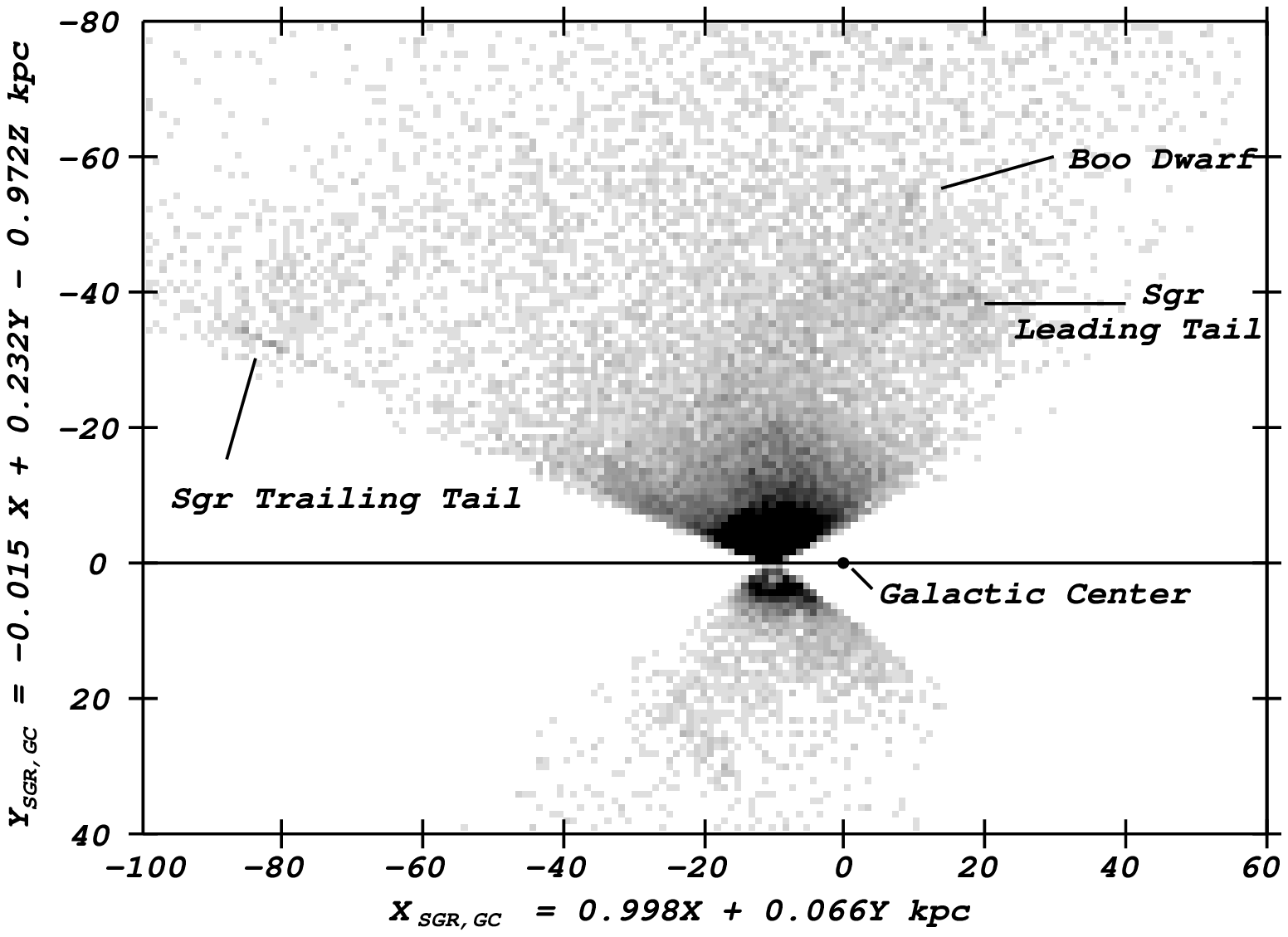}

%\plotone{lgbhb5n.eps}
\plotone{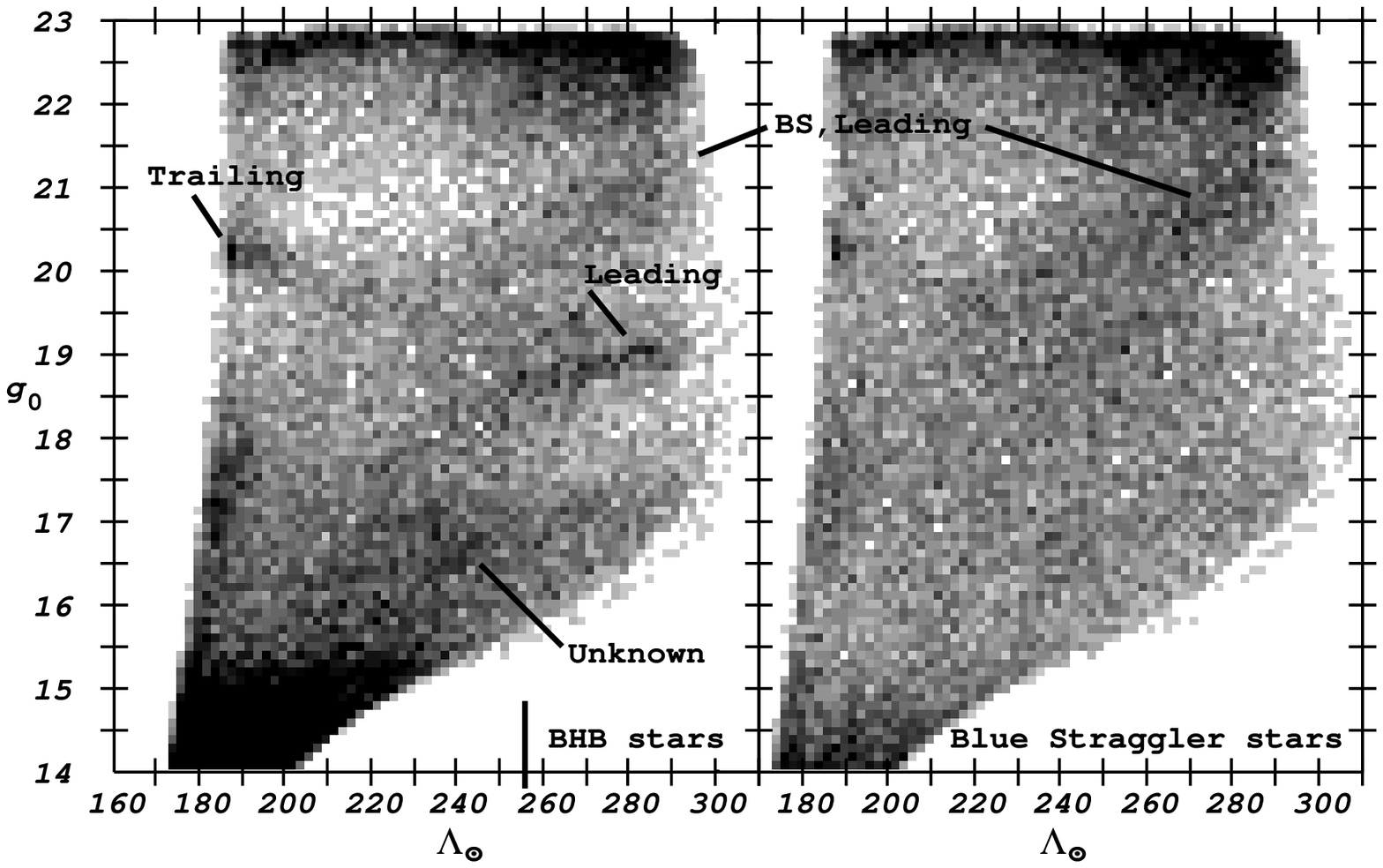}

\clearpage
%\plotone{lgbhbsidebyside14n.eps}
\plotone{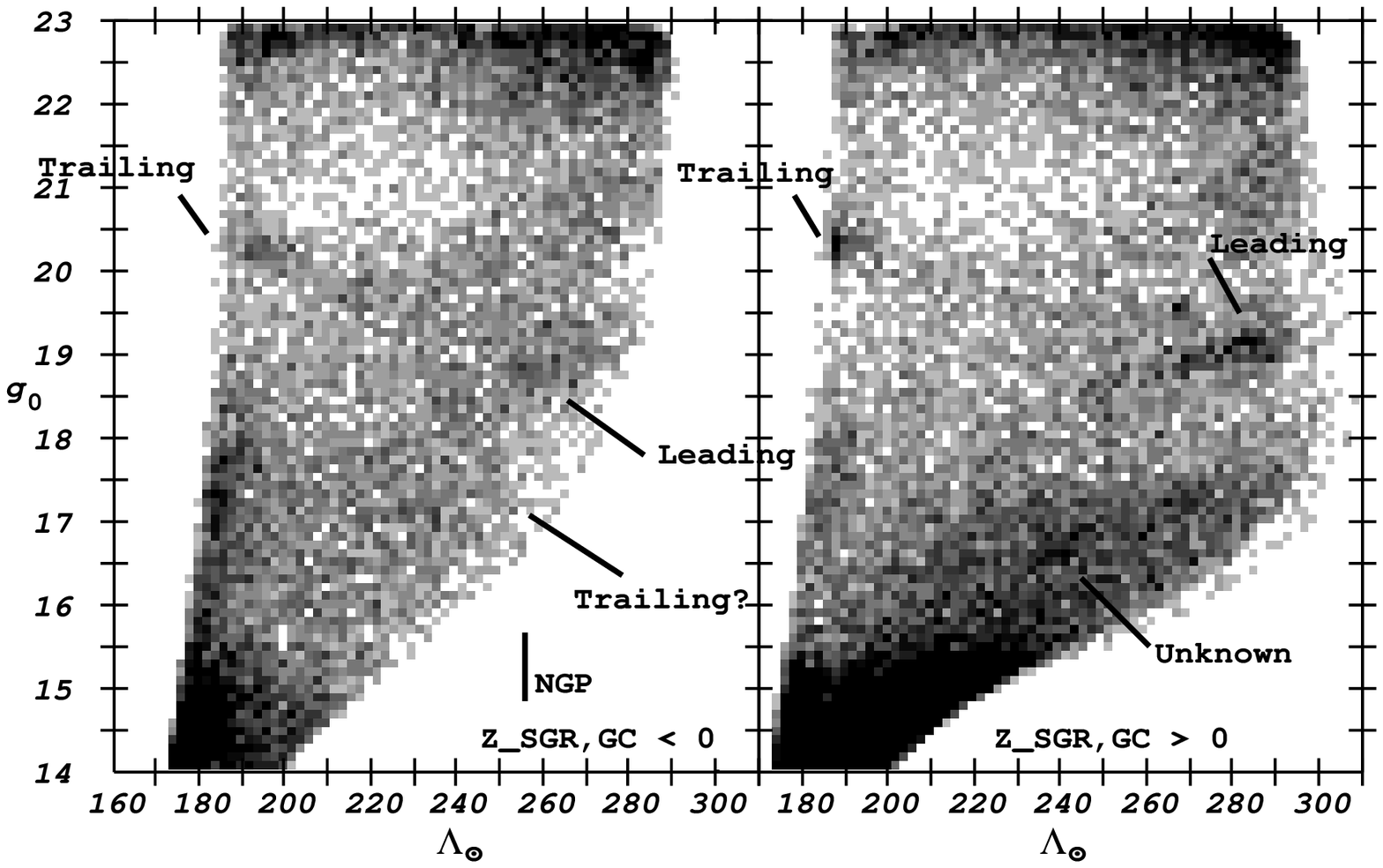}

%\plotone{f11n.eps}
\plotone{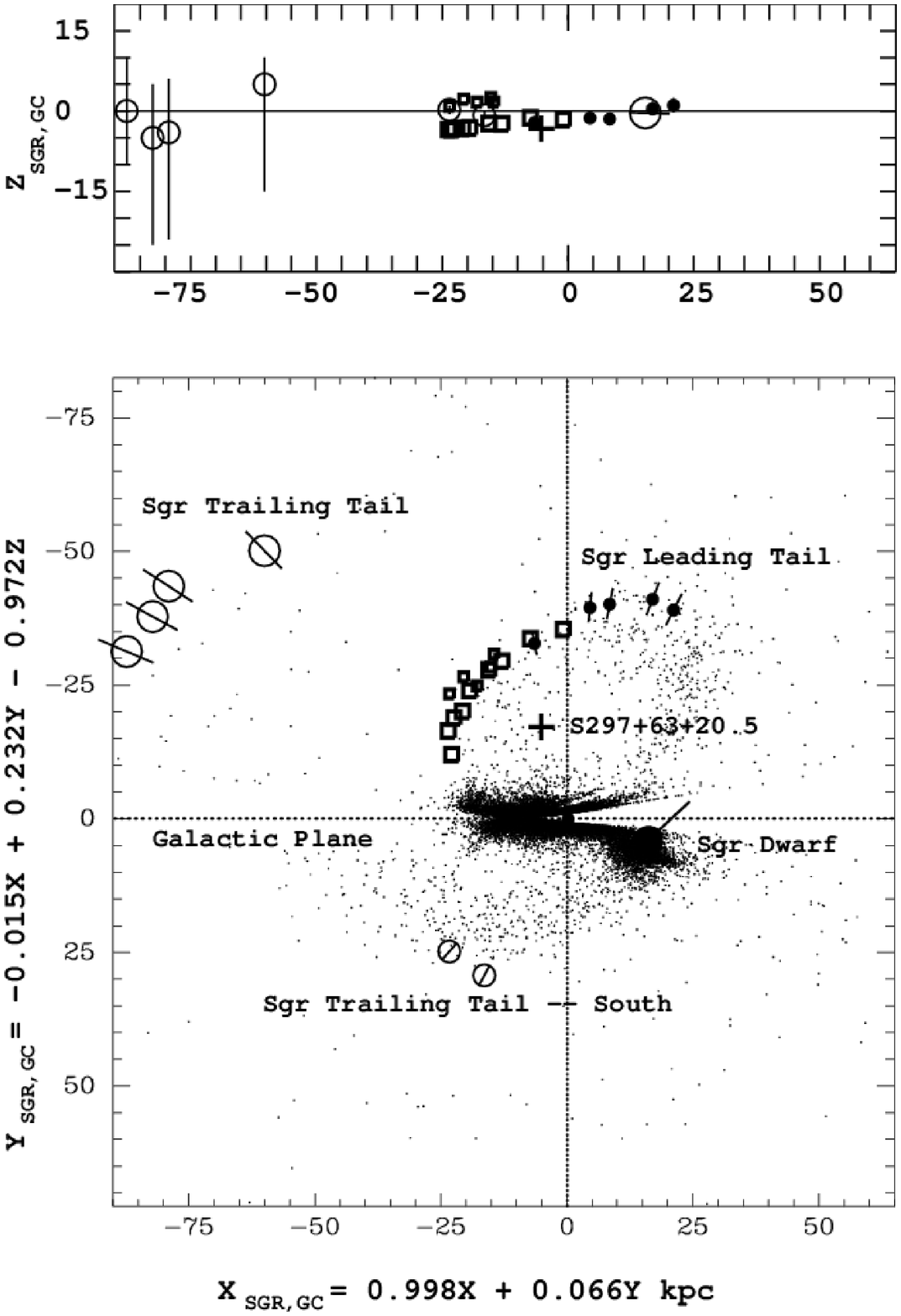}
%\plotone{f4hires.eps}

%\plotone{figpolarn.eps}
\plotone{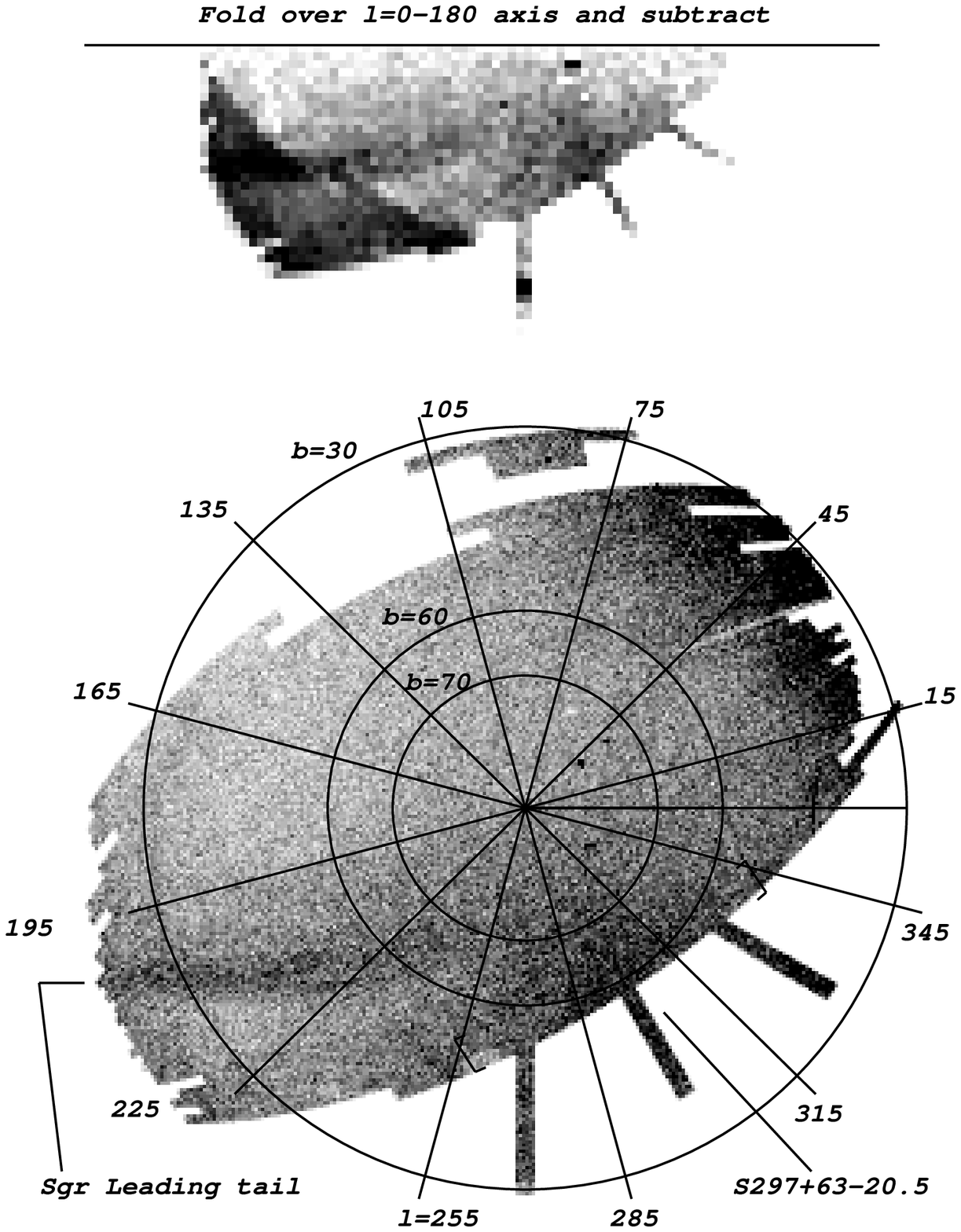}

%\plotone{figpolar2n.eps}
\plotone{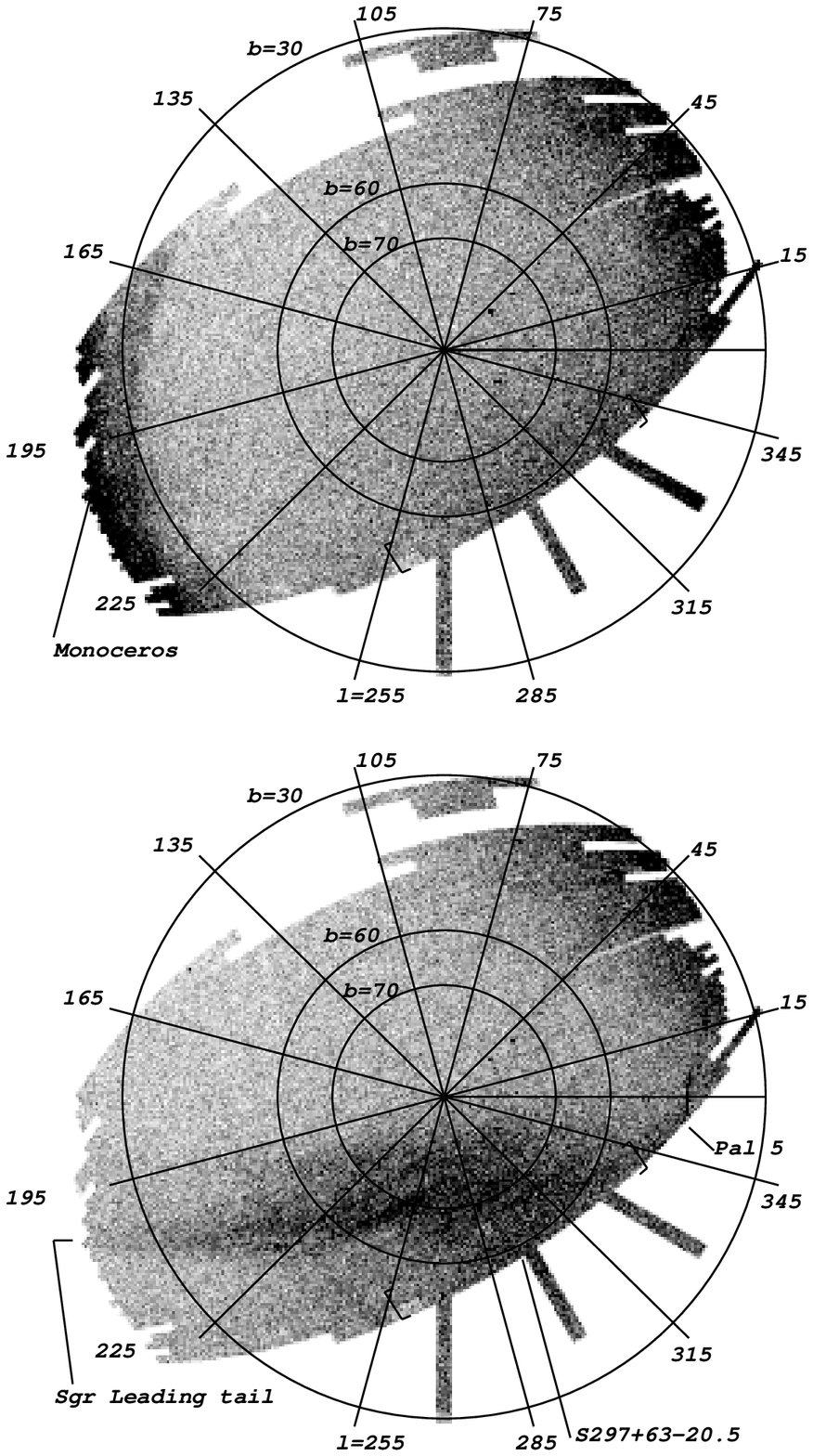}

%\plotone{fcountsonplatesn.eps}
\plotone{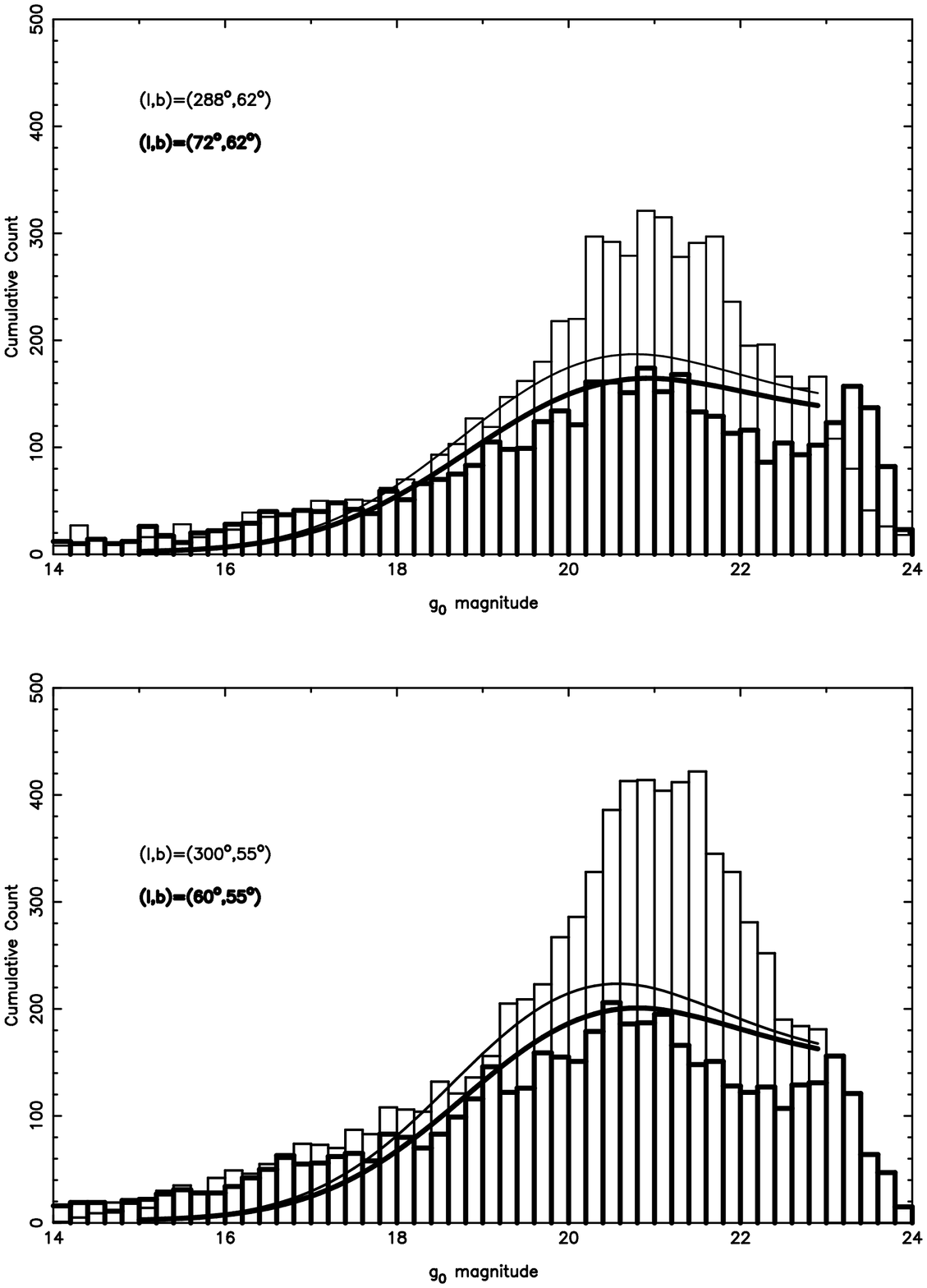}

%\plotone{gentypen.eps}
\plotone{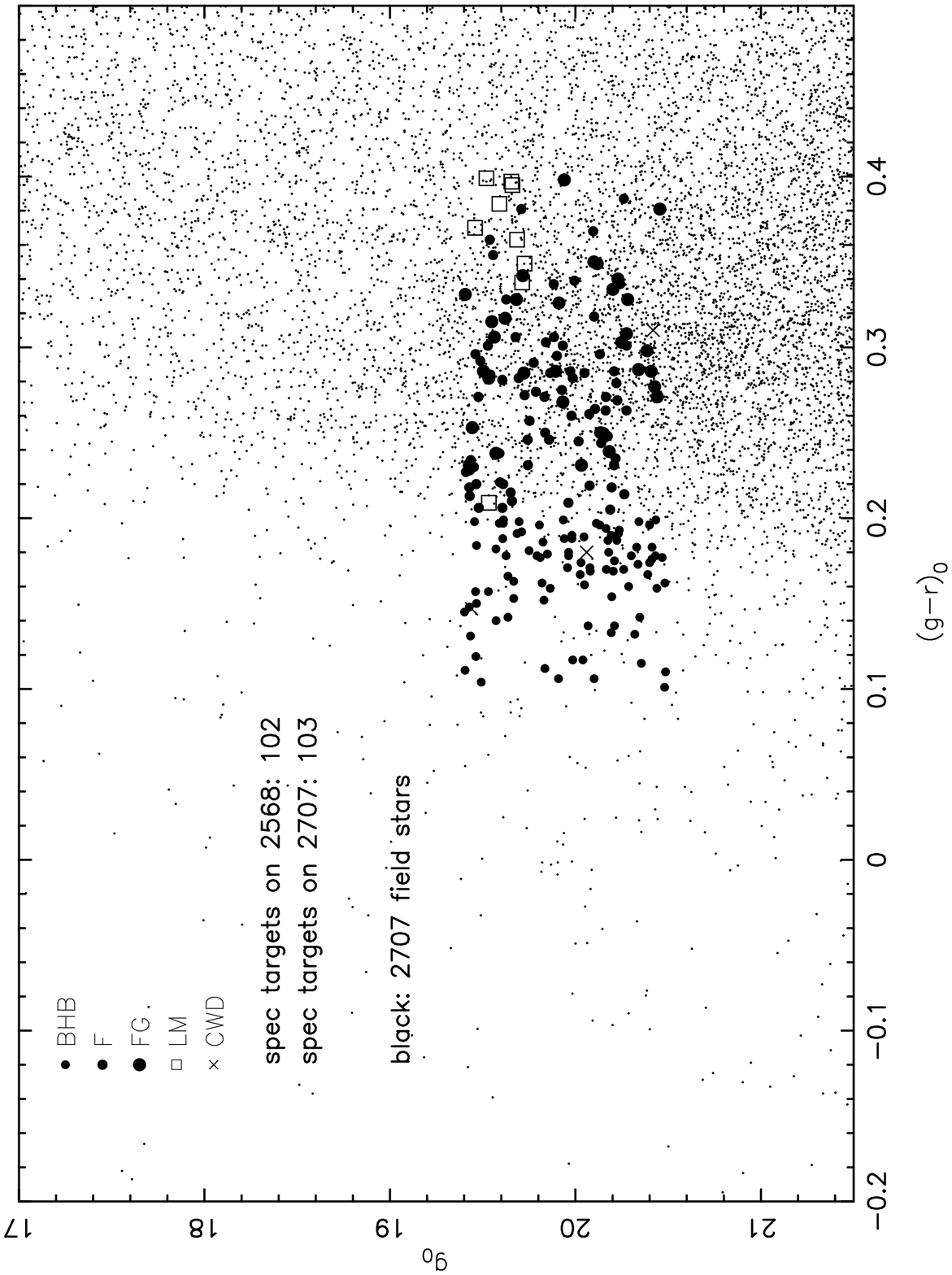}

%\plotone{virgorvn8n.eps}
\plotone{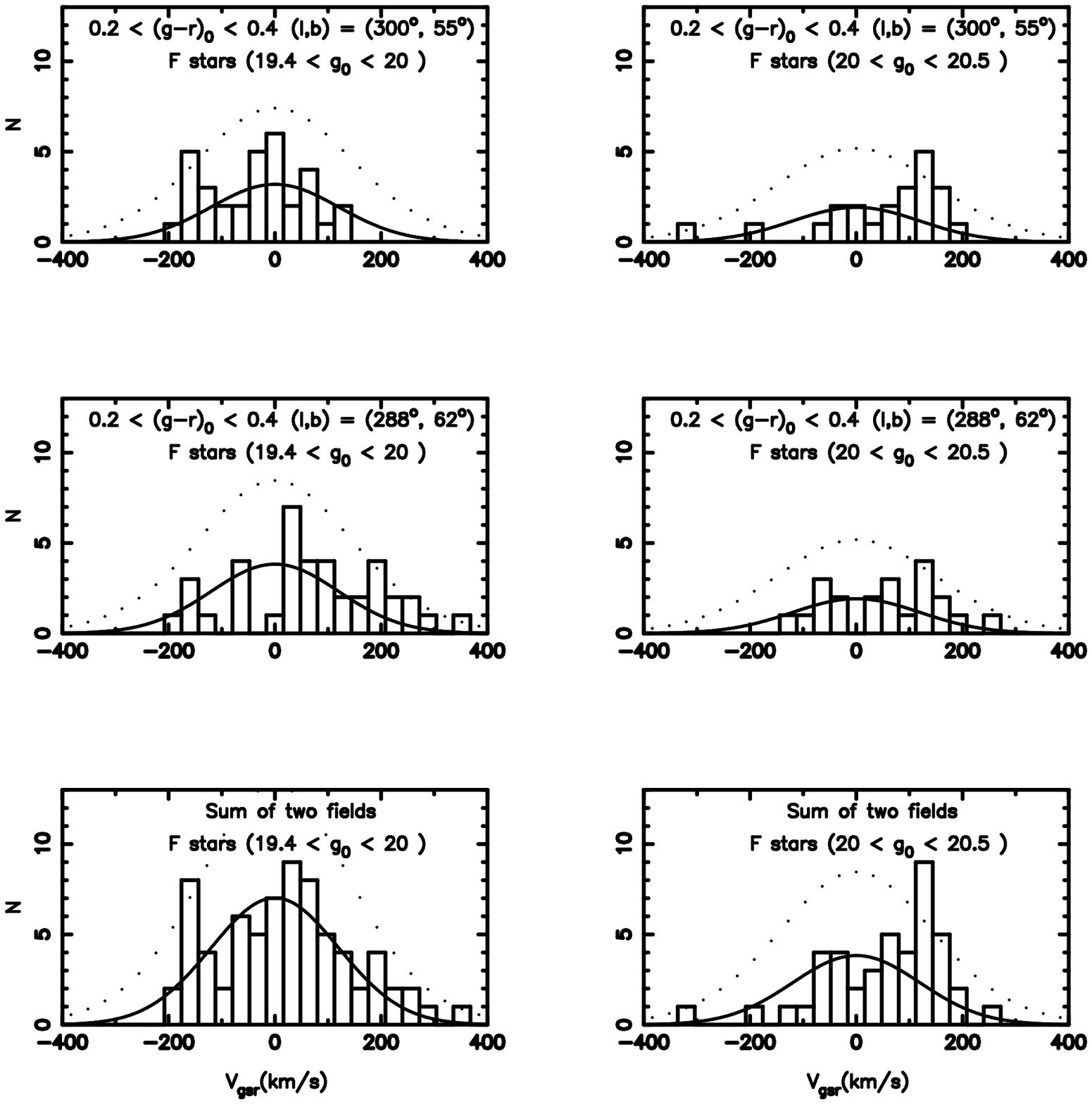}

%\plotone{virgorvn8n2.eps}
\plotone{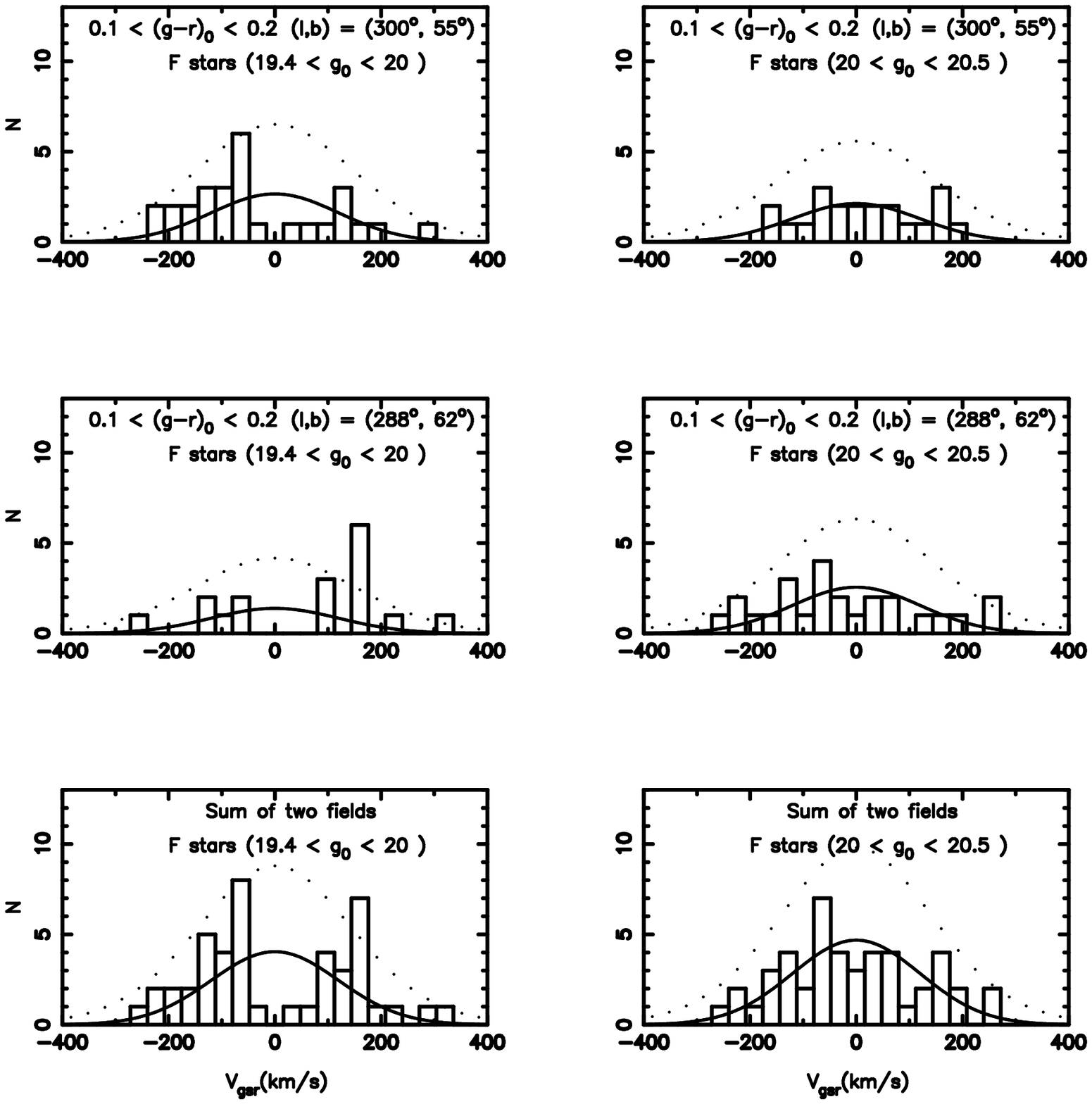}

\end{document}